\documentclass[preprint,12pt]{elsarticle}

\usepackage{amssymb}

\usepackage{graphicx}
\usepackage{color}
\usepackage{array}
\newcolumntype{C}[1]{>{\centering\let\newline\\\arraybackslash\hspace{0pt}}m{#1}}

\usepackage{subcaption}

\usepackage[usenames,dvipsnames]{xcolor}

\usepackage{tikz}
\usetikzlibrary{shapes,arrows}
\usetikzlibrary{decorations.markings}
\usetikzlibrary{decorations.pathreplacing}

\usepackage{amsmath}
\usepackage{amsfonts}

\usepackage{amsthm}
\usepackage{float}
\usepackage{changepage}

\usepackage{natbib}

\journal{}

\begin{document}

\begin{frontmatter}



\title{A Mathematical Model for Lymphangiogenesis \\in Normal and Diabetic Wounds}


\author[hwu]{Arianna~Bianchi}
\ead{ab584@hw.ac.uk}

\author[hwu]{Kevin~J.~Painter}

\author[hwu]{Jonathan~A.~Sherratt}

\address[hwu]{Department of Mathematics and Maxwell Institute for Mathematical Sciences,
	Heriot-Watt University, Edinburgh, Scotland, EH14 4AS, UK}

\begin{abstract}  
Several studies suggest that one possible cause of impaired wound healing is failed or insufficient lymphangiogenesis, that is the formation of new lymphatic capillaries. Although many mathematical models have been developed to describe the formation of blood capillaries (angiogenesis) very few have been proposed for the regeneration of the lymphatic network. 
Moreover, lymphangiogenesis is markedly distinct from angiogenesis, occurring at different times and in a different manner.
Here a model of five ordinary differential equations is presented to describe the formation of lymphatic capillaries following a skin wound. The variables represent different cell densities and growth factor concentrations, and where possible the parameters are estimated from experimental and clinical data.
The system is then solved numerically and the results are compared with the available biological literature.
Finally, a parameter sensitivity analysis of the model is taken as a starting point for suggesting new therapeutic approaches targeting the enhancement of lymphangiogenesis in diabetic wounds.
The work provides a deeper understanding of the phenomenon in question, clarifying the main factors involved. 
In particular, the balance between TGF-$\beta$ and VEGF levels, rather than their absolute values, is identified as crucial to effective lymphangiogenesis.
In addition, the results indicate lowering the macrophage-mediated activation of TGF-$\beta$ and increasing the basal lymphatic endothelial cell growth rate, \emph{inter alia}, as potential treatments.
It is hoped the findings of this paper may be considered in the development of future experiments investigating novel lymphangiogenic therapies.
\end{abstract}


\end{frontmatter}



\section{Introduction}

\subsection{Motivation}

Much effort has been spent in order to better understand and potentially treat the impairment of wound healing in diabetic patients. In this regard, one phenomenon that has recently gained
attention from biologists is \emph{lymphangiogenesis};
that is, the formation or reformation of lymphatic vasculature
\cite{huggenberger2011,huggenberger2011b,kim2012,tammela2010}. Insufficient lyphangiogenesis, as observed in diabetic subjects, appears to correlate with failed or delayed wound healing.

Impaired wound healing is a major health problem worldwide
and in recent decades has attracted the attention of both biologists and mathematicians.
In many cases unresolved wound healing correlates with prolonged infection,
which negatively affects the patient's quality of life, 
causing pain and impairing their physical abilities.
Particularly serious infection may even require the amputation of an affected limb 
\cite{langer2009}.
Furthermore, impaired wound healing also constitutes a major problem for health care systems,
accounting for approximately 3\% of all health service expenses in the UK  
and 20 billion dollars annually in the USA \cite{delatorre2013,dowsett2009,drew2007,posnett2008}.
Several systemic factors contribute 
to the delay or complete failure of the wound healing process
 \cite{asai2012,fadini2010,jeffcoate2003,lerman2003,swift2001}.
In particular, diabetic patients exhibit a slower and sometimes insufficient response to infection after injury.
Such a delay often results in a chronic wound; that is, the wound fails to progress through the normal stages of healing
and usually remains at the inflammation stage \cite{brem2007,pierce2001}.

Interest in lymphangiogenesis in reference to wound healing is very recent:
for example in the Singer \& Clark 1999 review \cite{singer1999} the process is not mentioned.
Nevertheless, lymphatic vessels have recently become regarded as a crucial factor in wound healing \cite{cho2006,ji2005}.
They mediate the immune response and maintain the right pressure in the tissues \cite{swartz2001},
thus playing a very important role in inflammation and contributing to the healing of a wound \cite{oliver2002,witte2001}.
Moreover, failed restoration of a lymphatic network (observed, for example, in diabetic patients)
is now thought to be a major cause of impairment to wound healing \cite{asai2012,maruyama2007,saaristo2006}.

Mathematical modelling has proven a useful tool 
in understanding the mechanisms behind numerous biological processes. 
It is therefore of interest, and potentially great utility, to build a model 
describing lymphangiogenesis in wound healing,
considering both the normal and pathological (diabetic) cases.

\subsection{Biology}

Wound healing is a very complex process involving a number of entwined events, 
which partially overlap in time and influence one another.
For simplicity and educational purposes,
it is often divided into four different phases:
hemostasis, inflammation, proliferation and remodeling.
Here the key events in each of the phases are summarised; for further details, 
see for instance \cite{delatorre2013,gabriel2013,park2013,singer1999,stadelmann1998}).
A few minutes after injury, the contact between blood and the 
extracellular matrix (ECM)
causes a biochemical reaction that leads to the formation of a blood clot.
This ``crust'' has the double function of stopping the bleeding (\emph{hemostasis})
and providing a ``scaffold'' for other cells involved in the process to be described below.
Concurrently, chemical regulators (such as Transforming Growth Factor $\beta$, or TGF-$\beta$) are released,
which attract cells such as neutrophils and monocytes to the wound site.
These cells clean the wound of debris and neutralise any infectious agents
that have invaded the tissue. 
This stage is called \emph{inflammation}; 
in a normal wound inflammation begins a couple of hours after wounding 
and lasts a few weeks.
Monocytes metamorphose into macrophages,
which complete the removal of the pathogens and also secrete some proteins 
(like Vascular Endothelial Growth Factor, or VEGF).
This leads to the next stage: the \emph{proliferation} or \emph{reepithelialisation} phase.
At this point VEGF and other substances stimulate the growth and aggregation of the surrounding cells,
restoring the different tissue functions.
The clot is slowly substituted by a ``temporary skin'' called \emph{granulation tissue}
and the interrupted blood and lymphatic capillary networks are restored in processes
named \emph{(blood) angiogenesis} and \emph{lymphangiogenesis}, respectively.
After a lengthy period the granulation tissue is replaced by normal skin;
this happens during the final, long phase of \emph{remodeling},
which can take up to one or two years.

Although wound healing has been studied extensively
and the main underlying mechanisms are well understood,
little is known about how lymphangiogenesis takes place.
Far more biological (and mathematical) literature has been produced 
about its sibling process, blood angiogenesis;
it was not until the 1990s that lymphangiogenesis received 
significant attention from researchers
\cite{adams2007,benest2008,choi2012}.
This discrepancy was mainly due to the previous lack of markers and information
on the growth factors involved in the lymphangiogenesis process;
such a dearth of biochemical tools impeded a detailed and quantifiable
study of lymphatic dynamics \cite{choi2012,oliver2002}. For biological reviews about lymphangiogenesis see \cite{kim2012,norrmen2011,tammela2010}
and for particularly significant biological research papers see \cite{boardman2003,bruyere2010,rutkowski2006}.

\begin{figure}[h]
     \centering
     \includegraphics[height=6cm]{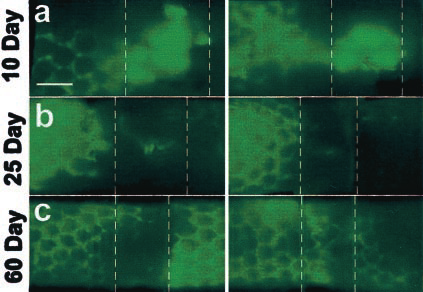}   
		\caption{The photo is taken from \cite[Figure 2]{boardman2003} and shows lymphatic channels formation in the mouse tail. Note that at day 10 (a) fluid channeling is not observed, but at day 25 (b) discrete channels are present and at day 60 (c) a hexagonal lymphatic network is nearly complete.
Notice also that lymph and interstitial fluid flows from left (tail tip) to right (tail base): this is in contrast to what happens during blood angiogenesis, which occurs equally from both sides of the implanted tissue equivalent. 
In 	\cite{boardman2003} the authors use a new model of skin regeneration consisting of a collagen implant in a mouse tail (whose location is indicated by the dashed lines in the picture). The aim of the experiment is mainly to characterise the process of lymphatic regeneration. 
Lymph fluid is detected (green in the photograph) and in \cite{boardman2003} it is shown that LECs follow this fluid. 
Therefore, this photo can be seen as the migration of LECs into the wound. (Bar=1mm)   }  
	\label{fig:fotoBoardman}
\end{figure}

Naively, lymphatic vessels may appear
``interchangeable'' with their blood equivalents from a modelling perspective.
However, it is stressed that the two vasculatures are quite different; for biological papers comparing lymphangiogenesis with (blood) angiogenesis see \cite{adams2007,lohela2009,sweat2012}.
First of all, the capillary structure is completely distinct:
while blood vessel walls are relatively thick, 
surrounded by smooth muscles which pump the blood around the body,
lymphatic capillaries are made of a single layer of 
endothelial cells known as \emph{lymphatic endothelial cells} (LEC) \cite{norrmen2011}.
Moreover, the formation of new lymphatic capillaries, 
or the restoration of preexisting ones, is very different from blood angiogenesis.
While growing blood capillaries are known to sprout from existing interrupted ones,
several studies suggest that lymphangiogenesis occurs in a different way \cite{benest2008,nakao2010}.
For instance, in \cite{rutkowski2006} it is observed that LECs migrate as single cells in the direction of interstitial flow and after sufficient numbers have congregated in the wound region, they organise into vessels (see Figure \ref{fig:fotoBoardman}).

We are therefore facing a new process whose mathematical description
cannot be drawn from any previous model for blood angiogenesis.
In the following section more details are given about the lymphangiogenesis process
and a mathematical model is proposed.

\subsection{Outline}
Clearly, successful lymphangiogenesis is an essential element in the wound healing process. Yet, as a novel and developing area of attention, its mechanistic basis, relation to other components of wound healing and impairment during diabetes remain unclear. With the aim of furthering our understanding of lymphangiogenesis, in this paper a mathematical model is developed to describe this process.

This paper is structured as follows.
In Section \ref{sec:modelling} a brief review of mathematical modelling of wound healing and lymphangiogenesis is given, followed by a detailed description of how the terms in our model were chosen. At the end of the section, a list of the model parameters and initial conditions is provided (a detailed description of parameter estimation can be found in \ref{appPAR}).
In Section \ref{sec:results} a typical solution of the model is shown and compared with real biological datasets.  Furthermore, the diabetic case is introduced and modelled by changing specific parameters. The simulations for normal and diabetic cases are then run together and compared with available data. Finally, a steady state analysis (detailed in \ref{appSS}) and parameter sensitivity analysis of the model are performed.
In Section \ref{sec:therapies} three existing experimental treatments aimed at enhancing lymphangiogenesis are presented and simulated; then, potential therapeutic targets are identified based on observations from the parameter sensitivity analysis.
Finally, Section \ref{sec:conclusions} briefly summarises the main results of the work and possible future extensions of the present model are mentioned.


\section{Modelling}  \label{sec:modelling}

\subsection{Brief review of existing models}
Being of such a complicated nature and evident medical interest,
wound healing has been the subject of mathematical modelling 
for decades; see for instance \cite{sherratt1990,tranquillo1992} for the first models in the 1990s, \cite{sherratt2002} for a 2002 review and \cite{geris2010} for a 2010 review.
A variety of mathematical formalisms have been involved in wound models:
from classical PDEs \cite{dale1995,olsen1997,sherratt1990}, 
often derived from bio-mechanical considerations \cite{friedman2011,murphy2012},
to stochastic models \cite{boyarsky1986}, to discrete models \cite{dallon1999}. 
Some authors have used moving boundary methods to study the movement
of the wound edge during healing \cite{javierre2009}, and attempts have been
made to understand more specific aspects of the healing process, for 
example macrophage dynamics in diabetic wounds \cite{waugh2006} and the resolution of the
inflammatory phase \cite{dunster2014}.
One aspect of wound healing that has received considerable attention from mathematicians is (blood) angiogenesis
(see, for instance, \cite{byrne2000,chaplain2000,flegg2012,gaffney2002,machado2011,pettet1996,schugart2008,zheng2013}).
A comprehensive review of mathematical models for vascular network formation can be found in Scianna et al's 2013 review \cite{scianna2013}.
Little work has been done with regards to modelling lymphangiogenesis and, indeed,  \cite{scianna2013} cites only a limited number of mathematical works concerning this topic. 
Of these a representative sample is given by \cite{friedman2005,roose2008}
which deal with tumor lymphangiogenesis 
and the collagen pre-patterning caused by interstitial fluid flow, respectively.
More specifically, \cite{roose2008} uses the physical theory of rubber materials 
to develop a model explaining the morphology of the lymphatic network in collagen gels,
following the experimental observations of \cite{boardman2003}. This is the only existent model for lymphangiogenesis in wound healing known to the authors.
A further brief review of lymphatic modelling can be found in \cite{margaris2012},
where the phenomenon is approached from an engineering perspective.

In summary, a small number of papers have considered modelling the lymphangiogenesis process in the context of tumors; the modelling of lymphangiogenesis in wound healing is confined to \cite{roose2008}, where two fourth order PDEs are used to describe the evolution of the collagen volume fraction and of the proton concentration in a collagen implant.
That work does not address the healing process as a whole which is the aim of the present paper. 
Here a simple model (comprised of a system of ODEs) is presented that provides an effective description of the main dynamics observed in wound healing lymphangiogenesis.

\subsection{Model development}
In the present model five time-dependent variables are considered:
two chemical concentrations (TGF-$\beta$ and VEGF) 
and three cell densities 
(macrophages, LECs and lymphatic capillaries).
Their interactions are described in Figure \ref{fig:modelscheme} and the formulation of the ODE model is based on the following set of processes
(the full system is given by the set of equations (\ref{eq:system})).
The initial (or pre-wounding) state is altered when latent TGF-$\beta$ is activated 
(thus becoming active TGF-$\beta$, denoted in the sequel by $T$) by macrophages and enzymes
released immediately after wounding.
This active form of TGF-$\beta$ attracts more macrophages ($M$) to the wound site, 
through \emph{chemotaxis}.
Macrophages in turn produce VEGF ($V$), a growth factor that chemoattracts and 
stimulates the proliferation of LECs ($L$). 
Note that LEC growth is also inhibited by TGF-$\beta$.
In the final stage of the process, LECs cluster in a network structure, 
 transdifferentiating into lymphatic capillaries ($C$).
This latter process happens spontaneously, although it is enhanced by VEGF.

\begin{center}

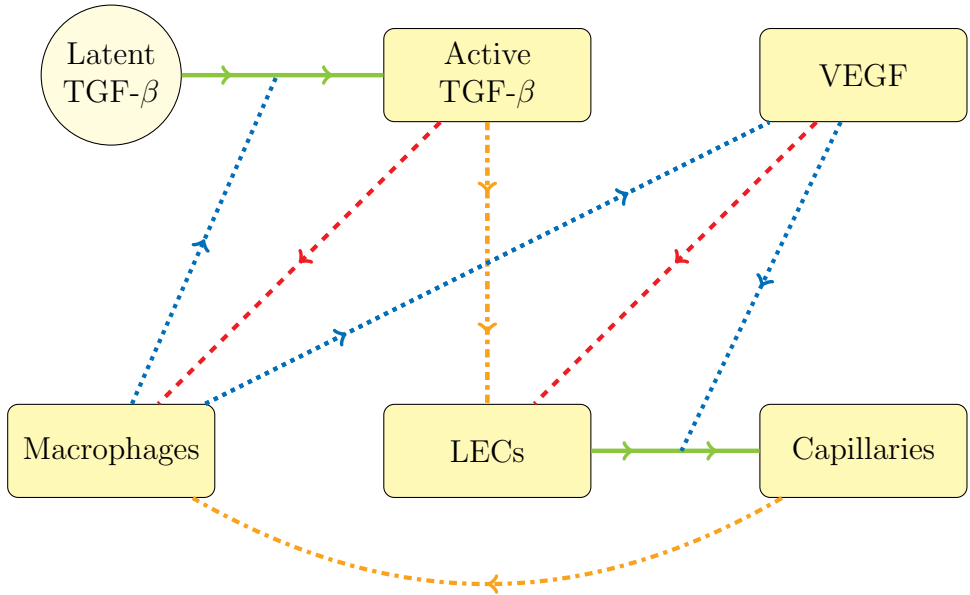
\begin{figure}[h]

\tikzstyle{constant} = [circle, draw, fill=yellow!15,text width=4em, text centered, node distance=2cm, inner sep=0pt]
\tikzstyle{block} = [rectangle, draw, fill=yellow!35,text width=6em, text centered, rounded corners, minimum height=3em]
\tikzstyle{line} = [draw, -latex']
\tikzset{->-/.style={decoration={
  markings,
  mark=at position .5 with {\arrow{>}}},postaction={decorate}}}
\tikzset{->--/.style={decoration={
  markings,
  mark=at position .25 with {\arrow{>}}}, postaction={decorate}}}
\tikzset{-->-/.style={decoration={
  markings,
  mark=at position .75 with {\arrow{>}}}, postaction={decorate}}}

\tikzstyle{solid}=[dash pattern=]
\tikzstyle{dashed}=[dash pattern=on 3pt off 3pt]
\tikzstyle{dotted}=[dash pattern=on \pgflinewidth off 2pt]
\tikzstyle{dashdotted}=[dash pattern=on 3pt off 2pt on \the\pgflinewidth off 2pt]

\label{modelscheme}

\begin{tikzpicture}[node distance = 5 cm, auto]  
\node [constant] (tl) {Latent TGF-$\beta$};
\node [block, right of=tl] (tgf) {Active TGF-$\beta$};
\node [block, right of=tgf] (vegf) {VEGF};
\node [block, below of=tl] (mph) {Macrophages};
\node [block, below of=tgf] (lec) {LECs};
\node [block, below of=vegf] (cap) {Capillaries};

\draw[ultra thick, solid,LimeGreen, ->--] (tl) -- node [midway, name=a1] {} (tgf);
\draw[ultra thick, solid,LimeGreen, -->-] (tl)  to (tgf);

\draw[ultra thick, dashed,Red, ->-] (tgf)  to  (mph);

\draw[ultra thick, dashdotted,YellowOrange, ->--] (tgf)  to  (lec);
\draw[ultra thick, dashdotted,YellowOrange, -->-] (tgf)  to  (lec);

\draw[ultra thick, dashed,Red,  ->-] (vegf)  to  (lec);

\draw[ultra thick, dotted,NavyBlue, ->--] (mph)  to  (vegf);
\draw[ultra thick, dotted,NavyBlue, -->-] (mph) to (vegf);

\draw[ultra thick, solid,LimeGreen, ->--] (lec) -- node [midway, below, name=a2] {} (cap);
\draw[ultra thick, solid,LimeGreen, -->-] (lec) to (cap);

\draw[ultra thick, dotted,NavyBlue, ->-] (mph) to (a1);
\draw[ultra thick, dotted,NavyBlue, ->-] (vegf) to (a2);

\draw[ultra thick, dashdotted,YellowOrange, ->-] (cap) to [bend left] (mph) ;
\end{tikzpicture}

\caption{A schematic representation of the model dynamics. The five time-dependent variables correspond to the levels of the factors in each of the rectangular boxes; the quantity in the circle (latent TGF-$\beta$) is assumed constant. Concerning the arrows, {\color{red} \textbf{dashed red}} denotes chemotactic attraction, {\color{LimeGreen} \textbf{solid green}} denotes activation/transdifferentiation, {\color{NavyBlue} \textbf{dotted blue}} denotes production/enhancement and {\color{YellowOrange} \textbf{dash-dotted orange}} denotes inhibition. }

\label{fig:modelscheme}

\end{figure}

\end{center}


In the following the time-dependent variables introduced above
are discussed in detail.
In particular, the derivation of the corresponding evolution equation is individually presented for each variable.


\subsubsection*{TGF-$\beta$} This chemical is normally stored 
in an inactive or latent form in the body; however, only \emph{active} TGF-$\beta$ plays an important role in wound healing lymphangiogenesis, and therefore we will only consider the dynamics of the active chemical.
Effectively, the active TGF-$\beta$ protein is bound to a molecule called Latency Associated Peptide,
forming the so-called Small Latent TGF-$\beta$ Complex. 
This in turn is linked to another protein called Latent TGF-$\beta$ Binding Protein,
overall forming the Large Latent TGF-$\beta$ Complex \cite{taylor2009}.
Hence, the two ``peptide shells'' must be removed before the organism can use the TGF-$\beta$.

Another feature of this growth factor is that it exists in at least three isoforms (TGF-$\beta$1, 2 and 3) which play different roles at different stages of wound healing \cite{cheifetz1990}. The isoform of primary interest in our application is TGF-$\beta$1.

The differential equation describing (active) TGF-$\beta$ concentration has the following form:

\begin{center}
    \begin{tabular}{ C{2.8cm} c C{2.5cm} c c c  c }
  change in TGF-$\beta$ concentration & $=$ &  activation by enzymes and macrophages &$\times$ & latent TGF-$\beta$ & $-$ & decay.    
    \end{tabular}
\end{center}

A review of the activation process is presented in \cite{taylor2009}
where it is reported that TGF-$\beta$ can be activated in two ways.
The first is \emph{enzyme-mediated activation} whereby enzymes, mainly plasmin, release the Large Latent TGF-$\beta$ Complex from the ECM and then the Latency Associated Peptide binds to surface receptors.
The second form of activation is \emph{receptor-mediated activation}. Here cells bind the Latency Associated Peptide and later deliver active TGF-$\beta$ to their own TGF-$\beta$-Receptors or to the receptors of another cell.
This behaviour is often observed in activated macrophages \cite{gosiewska1999,nunes1995}.

Thus both enzyme concentration and macrophage density $M$ are influential in the activation process and thereby appear in the activation term, $a_p p_0 e^{-a_p T_L t} + a_M M$.
Here $a_p$ and $a_M$ denote the activation rates by enzymes and by macrophages, respectively. $T_L$ denotes the (constant) amount of available latent TGF-$\beta$ (more details in the next paragraph).
In addition, the enzyme/plasmin concentration is assumed to decrease exponentially from the initial value $p_0$, as in \cite{dale1996}; this reproduces quite well the enzyme dynamics in real wounds \cite{sinclair1994}. 

It is widely accepted that a variety of cells have the potential to secrete latent TGF-$\beta$,
including platelets, keratinocytes, macrophages, lymphocytes and fibroblasts \cite{barrientos2008,khalil1993,taylor2009}.
Moreover, this latent complex is stored in the ECM in order to be constantly available to the surrounding cells \cite{shi2011}.
This fact is manifested in a constant production rate $T_L$ in our model equation.
Furthermore, it is well known that macrophages secrete latent TGF-$\beta$ \cite{khalil1993},
we assume that this occurs at a constant rate $r_1$.
Together these considerations imply that the amount of available latent TGF-$\beta$ in the wound will be modelled by $T_L + r_1 M$.
Finally, TGF-$\beta$ naturally decays at rate $d_1$,
so the term $- d_1 T$ will be included in the differential equation.
Therefore, the full equation for active TGF-$\beta$ is
\begin{equation}  \label{eq:Teqn}
\frac{dT}{dt} = \left[ a_p p_0\exp(-a_p T_L t) + a_M M \right]
                \cdot \left[ T_L + r_1 M \right] - d_1 T           \;  .
\end{equation}


\subsubsection*{Macrophages} These are a type of white blood cell 
that removes debris, pathogenic microorganisms and cancer cells through 
\emph{phagocytosis}.
They are produced by the differentiation of monocytes and are found in most of the tissues, 
patrolling for potential pathogens.

Perversely, in addition to enhancing inflammation and stimulating the immune system, macrophages can also contribute to decreased immune reactions. 
For this reason they are classified either
as \emph{M1} (or \emph{inflammatory}) macrophages if they encourage inflammation,
or as \emph{M2} (or \emph{repair}) macrophages if they decrease inflammation and encourage tissue repair \cite{martinez2006}.
Henceforth we restrict attention to inflammatory macrophages, since they are the most involved in lymphangiogenesis-related processes.
A useful review of the multifaceted and versatile role of macrophages in wound healing can be found in \cite{rodero2010}.

The following scheme will be considered for macrophage dynamics:

\begin{center}
    \begin{tabular}{ C{1.8cm} c C{1.6cm} c C{1.8cm} c C{1.5cm} c C{1.8cm} }
  \footnotesize  change in macrophage density & $=$ & \footnotesize  constant source & $+$ & \footnotesize  chemotaxis by TGF-$\beta$ & $+$ & \footnotesize  logistic growth & $-$ & \footnotesize  removal and metamorphoses. 
    \end{tabular}
\end{center}
The various terms appearing in the right-hand side of this equation are discussed below.

The number of inflammatory macrophages increases due to their migration into the wound, in part due to movement of existing inflammatory macrophages from the surrounding tissue, as well as by chemotaxis of monocytes up gradients of TGF-$\beta$ \cite{wahl1987}, a fraction $\alpha$  of which differentiate into inflammatory macrophages \cite{mantovani2004}.
The former is modelled by assuming a constant source $s_M$ dictated by the non-zero level of inflammatory macrophages \cite{weber1990},  and the latter by the term $\alpha  h_1 (T)  =  \alpha { b_1 T^2}/{(b_2 + T^4)}$.
Here $h_1(T)$ is the ``chemotactic function'', whose form is discussed in detail in \ref{appPAR}.
Only a (small) percentage $\beta$ of macrophages 
undergo mitosis~\cite{greenwood1973}; we thus assume the logistic growth term $\beta r_2 M \left( 1 - {M}/{k_1} \right)$ where $r_2$ denotes the macrophage growth rate and $k_1$ the carrying capacity of the wound.
Notice that here only $M$ appears over the carrying capacity and the other cell types $L$ and $C$ are omitted. However, since the logistic term is small overall, adding $L$ and $C$ here
would just increase the numerical complexity of the system 
without adding any significant contribution to the dynamics of the problem.  This is reflected in the parameter sensitivity analysis provided later in the paper, and simulations (not shown) including all populations showed no appreciable difference.
Finally, inflammatory macrophages can die, metamorphose into repair macrophages 
or be washed away by the lymph flow. This is embodied in the removal term $- ( d_2 + \rho C ) M$, where $d_2$ denotes the constant death rate. Here metamorphosis and removal are considered to be linearly proportional to the capillary density $C$ through the coefficient $\rho$: in particular, capillary formation is an index of progression through the healing process and, to reflect the decreased requirement for inflammatory macrophages as wounding proceeds, we assume the metamorphosis/removal rate increases with the size of $C$.
Combining these observations one derives the macrophage equation
\begin{equation}  \label{eq:Meqn}
\frac{dM}{dt} = s_M + \alpha \frac{ b_1 T^2}{b_2 + T^4} 
                + \beta r_2 M \left( 1-\frac{M}{k_1} \right) - (d_2 +  \rho C)  M   \; .
\end{equation}


\subsubsection*{VEGF} This is a signal protein 
whose main function is to induce the formation of vascular networks 
by stimulating proliferation, migration and self-organisation of cells
after binding to specific receptors on their surface.
There are many kinds of VEGF:
while VEGF-A and VEGF-B are involved mainly in blood angiogenesis, 
VEGF-C and VEGF-D are the most important biochemical mediators of lymphangiogenesis via the receptor VEGFR3
(although VEGF-C can also stimulate angiogenesis via VEGFR2).
For a comprehensive description of the growth factors involved in lymphangiogenesis see \cite{jussila2002,lohela2003}.
Henceforth ``VEGF'' refers to VEGF-C (and, to a lesser extent, VEGF-D), unless otherwise stated.
For VEGF we assume the following dynamics:

\begin{center}
    \begin{tabular}{ C{2cm} c C{1.6cm} c C{2.2cm} c c c C{1.5cm} }
  \footnotesize  change in VEGF concentration & $=$ & \footnotesize  constant source & $+$ & \footnotesize  production by macrophages & $-$ & \footnotesize  decay & $-$ & \footnotesize  use by LECs.
    \end{tabular}
\end{center}

\noindent
Since the normal VEGF level in the skin is nonzero \cite{hormbrey2003,papaioannou2009}, 
it is assumed there is a constant source $s_V$ of this growth factor from the surrounding tissues.
VEGF is produced by several cells, but macrophages are considered to be one of its main sources 
in the context of wound healing \cite{kiriakidis2003,xiong1998}.
It is therefore natural to add the production term $+ r_3 M$
to the VEGF equation, where $r_3$ is the production rate of the chemical by macrophages.
On the other hand, the VEGF level is reduced by natural decay at constant rate $d_3$,
taken into account by the term $- d_3 V $.
In addition VEGF is internalised by cells:
effectively, LECs use VEGF to divide and form capillaries \cite{matsumoto2001,zachary2001};
it is assumed this process occurs at a constant rate $\gamma$, leading to the term $- \gamma VL$.
Thus, in this model the equation for VEGF dynamics is
\begin{equation}  \label{eq:Veqn}
\frac{dV}{dt} = s_V + r_3 M - d_3 V -  \gamma V L  \;  .
\end{equation}


\subsubsection*{LECs} As discussed above, lymphatic vessel walls are made of (lymphatic) endothelial cells.
The equation describing the presence of LECs in the wound consists of the following terms:

\begin{center}
    \begin{tabular}{ C{1.5cm} c C{2.3cm} c C{1.8cm} c C{1.5cm} c C{1.6cm} }
  \footnotesize change in LEC density & $=$ &  \footnotesize  growth, upregulated by VEGF and downregulated by TGF-$\beta$ & $+$ &  \footnotesize  inflow and chemotaxis by VEGF & $-$ & \footnotesize crowding effect and apoptosis & $-$ &  \footnotesize  transdifferentiation into capillaries.
    \end{tabular}
\end{center}

\noindent
LEC growth is upregulated by VEGF \cite{bernatchez1999,whitehurst2006,zachary2001} 
and downregulated by TGF-$\beta$ \cite{muller1987,sutton1991}.
The former observation is described mathematically 
by augmenting the normal/basal constant growth rate $c_1$ with $V$ in an increasing saturating manner through the parameters $c_2$ and $c_3$.
To account for the latter, the growth term is multiplied by a decreasing function of $T$.
Explicitly, the whole proliferation term is
\begin{equation} \label{eq:Leq-growth}
 \left( c_1 + \frac{V}{c_2 + c_3 V} \right) \left( \frac{1}{1+c_4 T} \right) L 
\end{equation}
where $c_4$ takes into account the ``intensity'' of the TGF-$\beta$ inhibition on LEC growth.

\noindent
It is assumed that LECs are brought into the wound by lymph flow at a constant rate $s_L$
and are chemoattracted by VEGF \cite{bernatchez1999,tammela2010}.
Considering a chemotactic function $h_2(V)$ of the same form 
as that used for TGF-$\beta$-mediated chemotaxis (see \ref{appPAR}),
these phenomena are captured by the terms
$$
 s_L + h_2(V)   =   s_L + \frac{b_3 V^2}{b_4 + V^4}   \;  .
$$
Since both of these movements originate from the interrupted lymphatic vasculature at the edges of the wound,
this flow will tend to decrease as the lymphatic network is restored.
Hence, supposing a linear correlation between the term above and the lymphatic regeneration,
the former is multiplied by the piecewise linear function $f(C)$ defined by
\begin{equation}  \label{eq:f(C)def}
f(C) = \left\{ \begin{array}{cl}
               1 - \frac{C}{C^*}  &  \mbox{ if } C < C^*  \\
               0                  &  \mbox{ if } C \geq C^*
\end{array}  \right.                                            \;  .  
\end{equation}
Here $C^*$ is a capillary density threshold value above which the lymphatic network is functional and uninterrupted and LEC flow stops.
Hence the final term for LEC inflow and chemotaxis is
\begin{equation} \label{eq:Leq-chemotax}
   \left( s_L + \frac{b_3 V^2}{b_4 + V^4} \right) f(C)  \;  .
\end{equation}

LEC growth is limited by over-crowding of the wound space,
a fact that is taken into account by the negative term $- L{(M+L+C)}/{k_2}$ where $k_2$ relates to the carrying capacity.
Finally, individual or small clusters of LECs migrate into the wound and later form
multicellular groups that slowly connect to one another,
organising into vessel structures \cite{boardman2003,rutkowski2006}.
Here it is assumed that when LECs are sufficiently populous
(that is, their density becomes larger than a threshold value $L^*$) 
they self-organise into capillaries at a rate which depends linearly on VEGF concentration via the term $\delta_2 V$.
In particular, endothelial cells tend to form network structures spontaneously (at a constant rate $\delta_1$, say)  but the rate is increased by the presence of VEGF \cite{podgrabinska2002}.
These observations result in the transdifferentiation term $- \sigma(L,C)\cdot (\delta_1 + \delta_2 V)L$
where $\sigma(L,C)$ is the step function
\begin{equation} \label{eq:sigma(L,C)def}
\sigma(L,C) = \left\{ \begin{array}{cl}
               1   &  \mbox{ if } L+C \geq L^*  \\
               0   &  \mbox{ if } L+C   <  L^*
\end{array}  \right.                                 \;  .
\end{equation}
Note that $\sigma$ depends both on $L$ and $C$: this is justified by the observation that the self-organisation process begins when $L$ reaches the threshold $L^*$ and then continues as LECs start forming the first capillaries.
Therefore the complete LEC equation is
\begin{eqnarray}  \label{eq:Leqn}
\frac{dL}{dt} & = & \left( c_1 + \frac{V}{c_2+c_3V} \right) \left(\frac{1}{1+c_4T}\right) L 
                    + \left( s_L + \frac{b_3 V^2}{b_4 + V^4} \right)f(C) \nonumber \\
              &   & - \frac{M+L+C}{k_2}L -  \sigma(L,C)\cdot (\delta_1 + \delta_2 V) L.  
\end{eqnarray}


\subsubsection*{Lymphatic capillaries} We assume that the lymphatic capillaries form simply from the self-organisation of LECs into a network structure.
Thus the capillary formation term is just the transdifferentiation term from the LEC equation above
and the dynamics of $C$ is modelled by
\begin{equation}  \label{eq:Ceqn}
\frac{dC}{dt} = \sigma(L,C)\cdot (\delta_1 + \delta_2 V) L  \; .
\end{equation}


The full system of equations is therefore given by

\begin{eqnarray}  \label{eq:system}
\frac{dT}{dt} & = & \underbrace{ \left[ a_p p_0\exp(-a_p T_L t) + a_M M \right] }
                             _{\stackrel{ \mbox{\footnotesize \footnotesize activation by}}{ \mbox{\footnotesize enzymes \& M$\Phi$s}}}
                \cdot \underbrace{ \left[ T_L + r_1 M \right] }_{ \mbox{\footnotesize latent TGF-$\beta$}}
                - \underbrace{ d_1 T }_{  \mbox{\footnotesize decay} }  \nonumber \\								
\frac{dM}{dt} & = & \underbrace{s_M}_{ \stackrel{ \mbox{\footnotesize constant}}{ \mbox{\footnotesize source}} }
                + \underbrace{ \alpha \frac{ b_1 T^2}{b_2 + T^4}}
								             _{\stackrel{ \mbox{\footnotesize chemotaxis}}{ \mbox{\footnotesize by TGF-$\beta$}}} 
                + \underbrace{ \beta r_2 M \left( 1-\frac{M}{k_1} \right) }_{  \mbox{\footnotesize logistic growth} }
                - \underbrace{ (d_2 +  \rho C)  M }_{ \stackrel{ \mbox{\footnotesize removal and}}{ \mbox{\footnotesize metamorphoses}} } \nonumber \\		
\frac{dV}{dt} & = & \underbrace{s_V}_{ \stackrel{ \mbox{\footnotesize constant}}{ \mbox{\footnotesize source}} }
                + \underbrace{ r_3 M }_{\stackrel{ \mbox{\footnotesize production}}{ \mbox{\footnotesize by M$\Phi$s}}}
                - \underbrace{ d_3 V }_{ \mbox{\footnotesize decay}}
                - \underbrace{ \gamma VL }_{ \mbox{\footnotesize use by LECs}} \nonumber \\
\frac{dL}{dt} & = & \underbrace{ \left( c_1 + \frac{V}{c_2+c_3V} \right) \left(\frac{1}{1+c_4T}\right) L }
                           _{\stackrel{  \mbox{\footnotesize growth upregulated by VEGF} }{ \mbox{\footnotesize and downregulated by TGF-$\beta$}}}
           + \underbrace{ \left( s_L + \frac{b_3 V^2}{b_4 + V^4} \right)f(C) }
					              _{\stackrel{ \mbox{\footnotesize inflow and}}{ \mbox{\footnotesize chemotaxis by VEGF}}}  \\
              &   & - \underbrace{\frac{M+L+C}{k_2}L}_{ \stackrel{ \mbox{\footnotesize crowding effect}}{ \mbox{\footnotesize and apoptosis}} }
           - \underbrace{ \sigma(L,C)\cdot (\delta_1 + \delta_2 V) L  }_{\stackrel{ \mbox{\footnotesize transdifferentiation}}{ \mbox{\footnotesize into capillaries}}} \nonumber \\
\frac{dC}{dt} & = & \underbrace{ \sigma(L,C)\cdot (\delta_1 + \delta_2 V) L }_{ \mbox{\footnotesize transdifferentiation of LECs} } \nonumber
\end{eqnarray}
where $f(C)$ and $\sigma(L,C)$ are defined in (\ref{eq:f(C)def}) and (\ref{eq:sigma(L,C)def}), respectively.


\subsection{Parameters and initial conditions}

\subsubsection*{Parameters}
Table \ref{table:parameters} gives a full list of parameter values, their units and the sources used for their estimation in the normal (non-diabetic) case.
It is remarked that great care was put into assessing the parameter values, and of the 31 parameters listed in the table, 25 have been estimated from biological data.
A detailed description of the estimation of each parameter can be found in \ref{appPAR}.

\begin{table}[p]

\begin{small}

\makebox[\textwidth][c]{\begin{tabular}{cccc}
\hline
\textsc{parameter}  &  \textsc{value} &  \textsc{units}                                       &  \textsc{source}   \\
\hline
     $a_p$  & $2.9\times 10^{-2}$  & $\mbox{mm}^3\mbox{pg}^{-1}\mbox{day}^{-1}$  & \cite{decrescenzo2001} \\
     $p_0$  & $2.5\times 10^5$          & $\mbox{pg mm}^{-3}$                         & no data found  \\
     $a_M$  & 0.45            & $\mbox{mm}^3\mbox{cells}^{-1}\mbox{day}^{-1}$   & \cite{gosiewska1999,nunes1995} \\
     $T_L$  & 18             & $\mbox{pg mm}^{-3}$                             & (\cite{oi2004}) \\
     $r_1$  & $3\times 10^{-5}$    & $\mbox{pg cells}^{-1}\mbox{day}^{-1}$      & \cite{khalil1993}  \\
     $d_1$  & $5\times 10^2$ & $\mbox{day}^{-1}$                          & \cite{kaminska2005}  \\
\hline     
     $s_M$    & $5.42\times 10^2$   & $\mbox{cells mm}^{-3}\mbox{day}^{-1}$   & (\cite{weber1990})\\
    $\alpha$  & 0.5             & 1                                                     & \cite{waugh2006}  \\
     $b_1$    & $8\times 10^8$  & $\mbox{cells pg}^2(\mbox{mm}^3)^{-3}\mbox{day}^{-1}$  & (\cite{nor2005})\\
     $b_2$    & $8.1\times 10^9$         & $(\mbox{pg mm}^{-3})^4$                           & \cite{wahl1987,yang1999} \\
     $\beta$  & $5\times 10^{-3}$           & 1                                                     & \cite{greenwood1973}  \\
     $r_2$    & 1.22          & $\mbox{day}^{-1}$                                     & \cite{zhuang1997}  \\
     $k_1$    & $6\times 10^5$& $\mbox{mm}^{3}\mbox{cells}^{-1}$               & \cite{zhuang1997}  \\
     $d_2$    & 0.2             & $\mbox{day}^{-1}$                             & \cite{cobbold2000} \\
     $\rho$   & $10^{-5}$     & $\mbox{day}^{-1}\mbox{cells}^{-1}$            & \cite{rutkowski2006} \\
\hline     
     $s_V$  & 1.9  & $\mbox{cells}\mbox{ day}^{-1}$ & (\cite{hormbrey2003,papaioannou2009})  \\
     $r_3$  & $1.9\times 10^{-3}$ & $\mbox{pg cells}^{-1}\mbox{day}^{-1}$  & (\cite{kiriakidis2003,sheikh2000})\\
     $d_3$  & 11                & $\mbox{day}^{-1}$                              & \cite{kleinheinz2010}  \\
   $\gamma$ & $1.4\times 10^{-3}$ & $\mbox{mm}^{3}\mbox{cells}^{-1}\mbox{day}^{-1}$& \cite{gabhann2004} \\
\hline     
     $c_1$          & 0.42         & $\mbox{day}^{-1}$                                     & \cite{nguyen2007}  \\
     $c_2$          & 42         & $\mbox{day}$                                          & \cite{whitehurst2006} \\
     $c_3$          & 4.1        & $\mbox{pg day mm}^{-3}$                               & \cite{whitehurst2006} \\
     $c_4$          & 0.24       & $\mbox{mm}^{3}\mbox{pg}^{-1}$                         & \cite{muller1987}\\
     $s_L$          & $5\times 10^2$       & $\mbox{cells day}^{-1}$                               & no data found \\
     $b_3$          & $10^7$    & $\mbox{cells pg}^2(\mbox{mm}^3)^{-3}\mbox{day}^{-1}$  & no data found \\
     $b_4$          & $8.1\times 10^9$   & $(\mbox{pg mm}^{-3})^4$                               & estimated $\approx b_2$\\
     $C^*$          & $10^4$          & $\mbox{cells mm}^{-3}$                                & \cite{rutkowski2006} \\
     $k_2$          & $4.71\times 10^5$    & $\mbox{cells day mm}^{-3}$                     & \cite{nguyen2007}\\
     $L^*$          & $10^4$          & $\mbox{cells mm}^{-3}$                                & \cite{rutkowski2006} \\ 
     $\delta_1$     & $5\times 10^{-2}$ & $\mbox{day}^{-1}$                                     & no data found \\
     $\delta_2$     & $10^{-3}$         & $\mbox{mm}^3\mbox{pg}^{-1}\mbox{day}^{-1}$            & no data found \\
\hline
\end{tabular}}

\end{small}

\caption{ A list of all the parameters appearing in the model equations (details of the estimation are provided in \ref{appPAR}). Each one is supplied with its estimated value, units and source used (when possible) to assess it. References in brackets mean that although the parameter was not \emph{directly} estimated from a dataset, its calculated value was compared with the biological literature; the caption ``no data found'' signifies that no suitable data were found to estimate the parameter. Concerning the VEGF value corresponding to maximum LEC chemotaxis $b_4$, it was assumed that its value is similar to its TGF-$\beta$ correspondent $b_2$; this choice was dictated by the lack of relevant/applicable biological data, to the authors' knowledge. }

\label{table:parameters}

\end{table}

\subsubsection*{Initial conditions}
In the present model, the initial time-point $t=0$ corresponds to the release of enzymes by platelets within the first hour after wounding \cite{sinclair1994,singer1999}.
The initial amounts of active TGF-$\beta$, macrophages and VEGF are taken to be their equilibrium values, 
estimated from experimental data as shown in Table \ref{table:initialconds}.
It is assumed that there are no endothelial cells or capillaries at $t=0$.

\begin{table}[h]

\begin{small}

\makebox[\textwidth][c]{\begin{tabular}{cccc}
\hline
\textsc{init.value}  &  \textsc{value} &  \textsc{units}  &  \textsc{source}   \\
\hline
     $T(0)$          &       30        & pg/mm$^3$        & \cite{yang1999} \\
     $M(0)$          &     1875        & cells/mm$^3$     & \cite{weber1990} \\
     $V(0)$          &        0.5      & pg/mm$^3$        & \cite{hormbrey2003,papaioannou2009} \\
     $L(0)$          &        0        & cells/mm$^3$     & assumption \\
     $C(0)$          &        0        & cells/mm$^3$     & assumption  \\
\hline
\end{tabular}}

\end{small}

\caption{ Values of the model variables at $t=0$. }

\label{table:initialconds}

\end{table}


\section{Results and analysis} \label{sec:results}

We now present a typical solution of the system (\ref{eq:system}) 
and compare it with biological data.
The system is solved numerically with the MatLab standard ODE solver \texttt{ode45} with relative tolerance $10^{-6}$ and absolute tolerance $10^{-9}$ over a time interval of 100 days.
It is remarked that the present model chiefly addresses inflammation and the early proliferation stage of the wound healing process.
In healthy subjects the inflammatory phase starts a few hours after injury and lasts approximately 1 or 2 weeks,
but it is prolonged in diabetic patients.
Moreover, lymphangiogenesis occurs between 25 and 60 days after wounding,
much later than blood angiogenesis which is observed between day 7 and day 17 \cite{benest2008,rutkowski2006}.
Thus, the equations are expected to realistically describe the phenomenon for about the first 100 days post-wounding.

\begin{figure}[h]
      \centering
      \makebox[\textwidth][c]{\includegraphics[height=4.5cm]{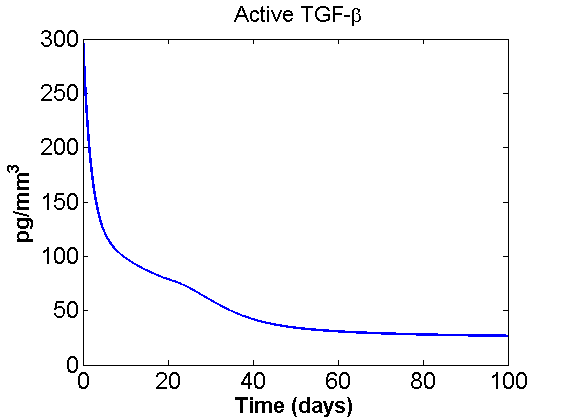}\includegraphics[height=4.5cm]{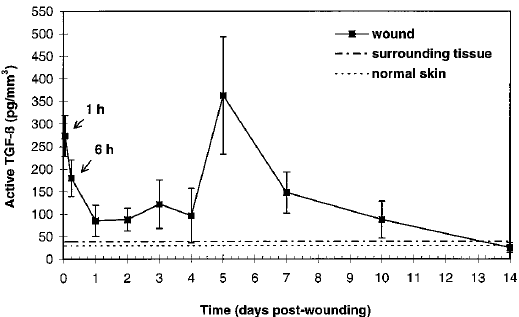}}
      
\caption{A comparison of the simulation output for the time course of TGF-$\beta$ concentration
with data from \cite[Figure 2]{yang1999}, showing the time course of active TGF-$\beta$ generation during wound repair in rats.}
      \label{fig:typicalsol-T}
\end{figure}

The TGF-$\beta$ level is expected to display a rapid spike in the first day post injury before returning to its equilibrium value \cite{yang1999}. 
In Figure \ref{fig:typicalsol-T} the simulation output is compared with a biological dataset. Both demonstrate the expected initial spike, but in the data a second peak is visible around day 5, reported also in \cite{nor2005}. We recall that TGF-$\beta$ exists in at least three known isoforms: TGF-$\beta$1, TGF-$\beta$2 and TGF-$\beta$3; the biological data set concerns all kinds of TGF-$\beta$ involved in other wound healing processes, such as collagen deposition, which are not modelled here (the time dynamics of the different TGF-$\beta$ isoforms can be found in \cite{yamano2013}). Nevertheless, the overall predicted trend of TGF-$\beta$ concentration in the wound matches the biological reality fairly well.

\begin{figure}[h]  
      \centering
      \includegraphics[height=4.5cm]{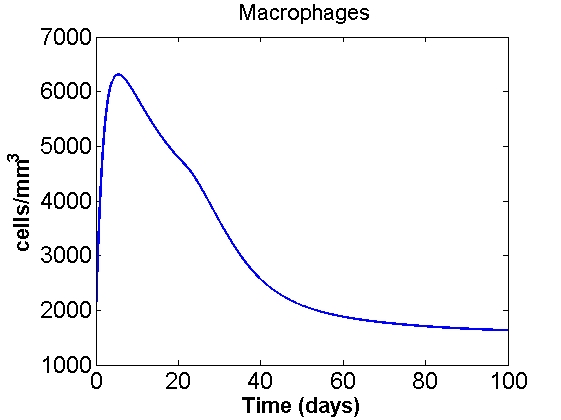}
      \includegraphics[height=4.5cm]{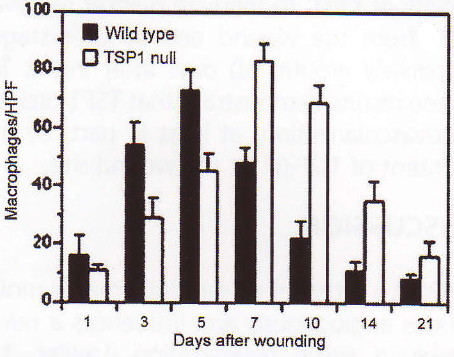}
      \caption{ A comparison between model and experimental data for the time course of macrophage density. Experimental data taken from \cite[Figure 3c]{nor2005}: note that the time-course comparison here is against the black bars, representing macrophage numbers in normal (wild-type) mice. }
      \label{fig:typicalsol-M}
\end{figure}

Macrophage levels are observed to reach a peak approximately 5 days after injury before returning to their equilibrium level \cite{nor2005}.
Again the model prediction is consistent with the biological literature,
as can be seen in Figure~\ref{fig:typicalsol-M}.

\begin{figure}[h]
      \centering
      \includegraphics[height=4.5cm]{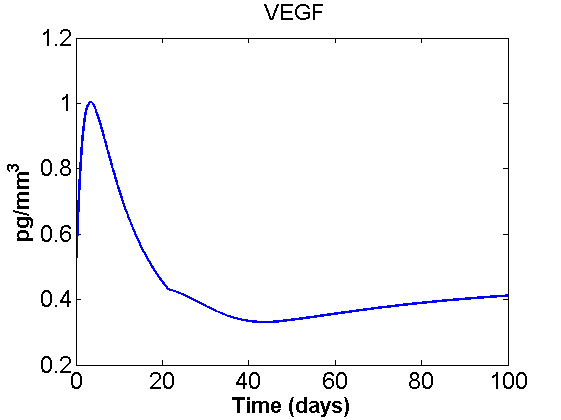}
      \includegraphics[height=5.5cm]{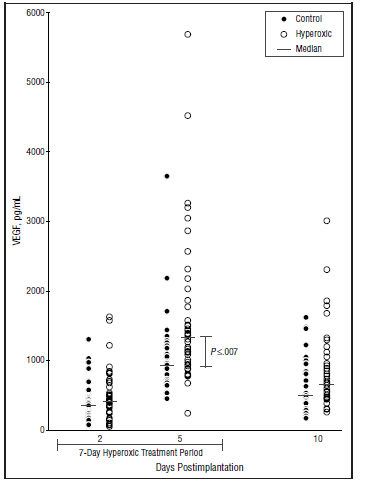} 
      \caption{A comparison of the simulation output for the time course of VEGF concentration with data from \cite[Figure 2]{sheikh2000}, where VEGF was measured in rat wound fluid (note that the units on the vertical axis are pg/mL, where 1000 pg/mL = 1 pg/mm$^3$).   }
      \label{fig:typicalsol-V}
\end{figure}

VEGF is also reported to reach its maximum concentration 5 days after wounding \cite{sheikh2000}. This is unsurprising given the above macrophage dynamics and the fact that macrophages are understood to be primarily responsible for the production of the protein.
Once again there is a strong correlation between the results of the theoretical model and experimental observations, as shown in Figure \ref{fig:typicalsol-V}.

\begin{figure}[h]  
      \centering
      \includegraphics[height=4.1cm]{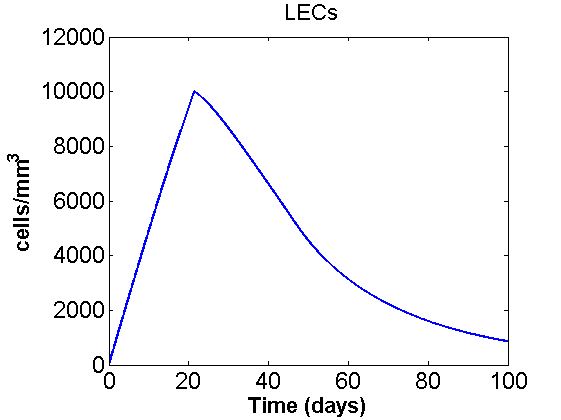}
      \includegraphics[height=4.1cm]{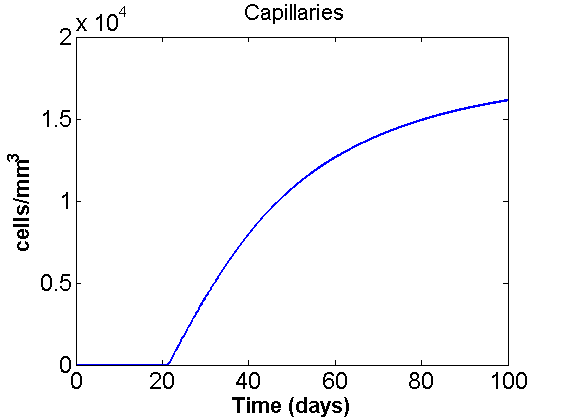}
			\includegraphics[height=4.1cm]{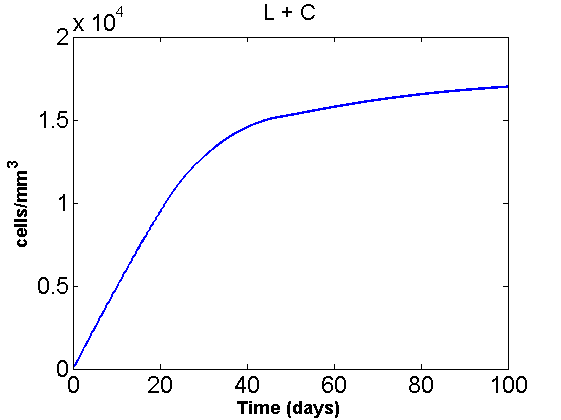}
			\caption{Simulation output for the time course of LEC density, lymphatic capillary density and their sum.
			         Note that the sum density has also been plotted the sum, 
			         since LEC and capillary cells are difficult to differentiate 
			         and any cell counts are likely to reflect the total density of these two cell types.}
      \label{fig:typicalsol-L+C}
\end{figure}

LEC levels are expected to increase immediately after wounding
but only later do the LECs self-organise into capillaries, around day 25 \cite{rutkowski2006}.
This is reflected in the simulation shown in Figure \ref{fig:typicalsol-L+C}.
Here LECs proliferate in the wound space until reaching the threshold level $L^* = 10^4$ around day 20. They then start agglomerating in capillary structures, commencing the lymphangiogenesis process proper.

\subsection{Modelling the diabetic case}  In order to simulate the diabetic case, some parameter values are changed as described in the following.
Unfortunately, it is difficult to obtain precise quantitative assessment of the appropriate changes,
and therefore the values chosen here only have a qualitative significance.

Several studies report that the TGF-$\beta$ level is significantly lower in diabetic wounds compared with controls.
This seems to be caused by impaired TGF-$\beta$ activation both by platelets and macrophages and by reduced production of TGF-$\beta$ by macrophages \cite{almulla2011,mirza2011,yamano2013}.
These features of the diabetic case are modelled by applying the following modifications to the parameters:
$$
a_p^{diab} = \frac{1}{2} a_p^{norm} < a_p^{norm} \quad , \quad a_M^{diab} = \frac{1}{2} a_M^{norm} < a_M^{norm}  \quad .  
$$
Furthermore, in diabetic wounds the macrophage density is higher than normal.
In particular, the inflammatory macrophage phenotype persists through several days after injury, showing an impaired transition to the repair phenotype \cite{mirza2011}.
In addition, macrophage functions (such as phagocytosis and migration) are impaired in the diabetic case \cite{khanna2010,xu2013}.
These differences from the normal case are reflected in the following choice of parameter changes:
\begin{equation*}
\begin{array}{r c l c r c l}
\alpha^{diab} &=& 0.8 > \alpha^{norm}, &\quad& b_1^{diab} &=& \frac{3}{4} b_1^{norm} < b_1^{norm}, \\ 
& & & & & & \\
d_2^{diab} &=& \frac{1}{2} d_2^{norm} < d_2^{norm},  &\quad&  r_3^{diab} &=& \frac{1}{2} r_3^{norm} < r_3^{norm}.
\end{array}
\end{equation*}

Finally, it is reported that endothelial cell proliferation 
is markedly reduced in diabetic wounds when compared with the normal case \cite{darby1997,curcio1992,kolluru2012}
(a detailed discussion of endothelial dysfunction in diabetes can be found in \cite{calles2001}).
This phenomenon is reflected in the model by reducing the basal proliferation rate of endothelial cells:
$$
c_1^{diab} = \frac{1}{2} c_1^{norm} < c_1^{norm} \; .
$$


\subsection{Comparison of results in the normal and diabetic cases}
Figures \ref{fig:sim-diab-T}-\ref{fig:sim-diab-L+C} show numerical simulations of the model comparing the time-course of the five model variables in the normal (blue solid line) and diabetic (red dashed line) case.

\subsubsection*{TGF-$\beta$} 
\begin{figure}[h]  
      \centering
      \includegraphics[height=4.5cm]{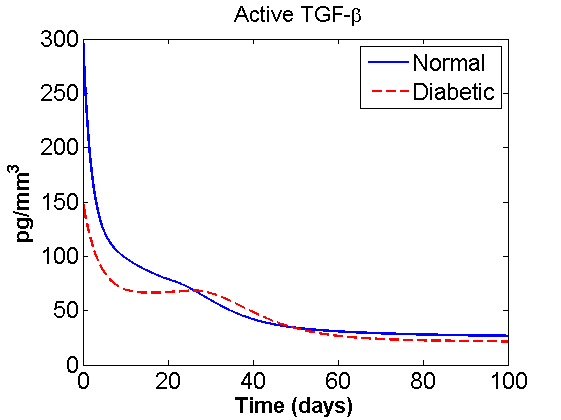}
\includegraphics[height=3.2cm]{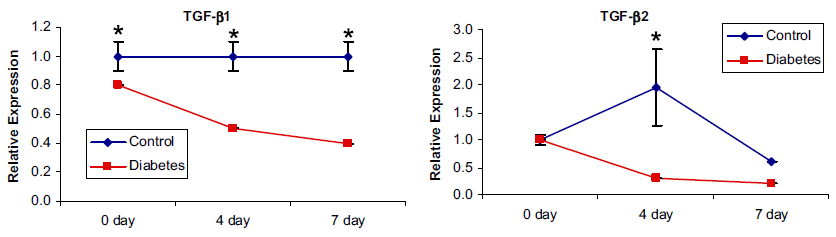}  
      \caption{ Time course of TGF-$\beta$ concentration in normal and diabetic wounds 
      as predicted by the model and as found empirically in  
      \cite[Figures 3G and 3H]{yamano2013}, where the authors study molecular dynamics during oral wound healing in normal (blue lines) and diabetic (red lines) mice.  }
      \label{fig:sim-diab-T}
\end{figure}

The level of TGF-$\beta$ recorded in diabetic wounds is lower than in the normal case,
at least in the first 20 days after injury \cite{almulla2011,mirza2011,yamano2013}. 
Model simulations are consistent with this (Figure \ref{fig:sim-diab-T}).

\subsubsection*{Macrophage}  
\begin{figure}[h]  
      \centering
      \makebox[\textwidth][c]{\includegraphics[height=4.5cm]{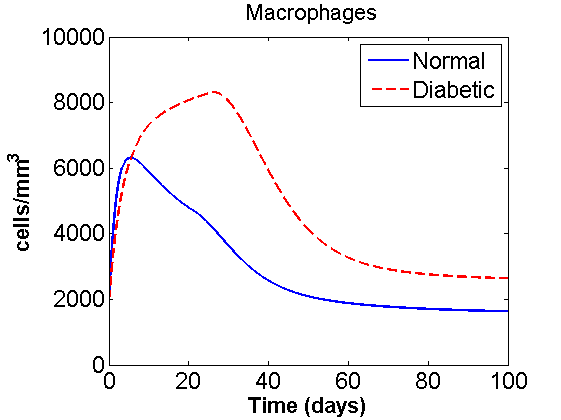}
			\includegraphics[height=4.5cm]{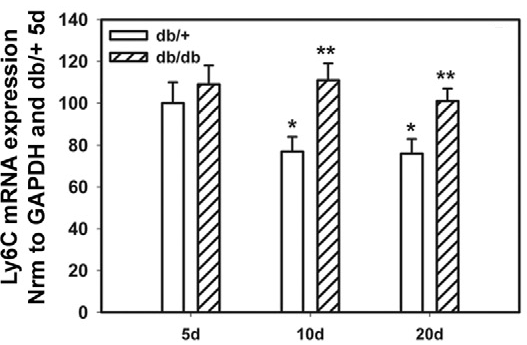}}     
      \caption{Time course of macrophage density in the normal and diabetic cases.
	  We compare the model prediction with \cite[Figure 1B]{mirza2011}.
In the experimental results, the relative height of shaded to solid white bars indicates the relative macrophage density in diabetic/non-diabetic wounds in mice (assessed via Ly6C expression, a marker for the macrophage lineage). }
      \label{fig:sim-diab-M}
\end{figure}

Experiments show that the density of macrophages in diabetic wounds is higher than in the normal case
and they persist for longer in the wound site \cite{mirza2011,rodero2010,xu2013}. Model simulations match these observations (Figure \ref{fig:sim-diab-M}).

\subsubsection*{VEGF} 
\begin{figure}[h]  
      \centering
      \includegraphics[height=4.5cm]{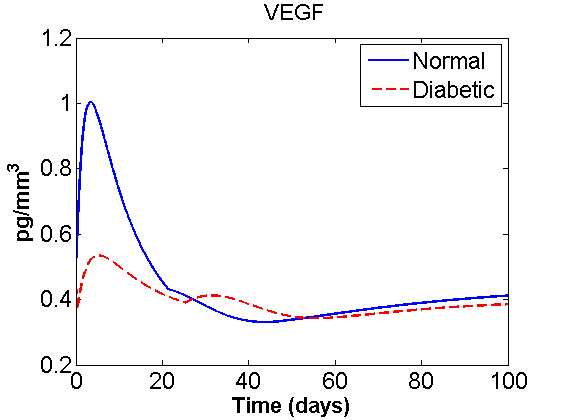}
			\includegraphics[height=4.5cm]{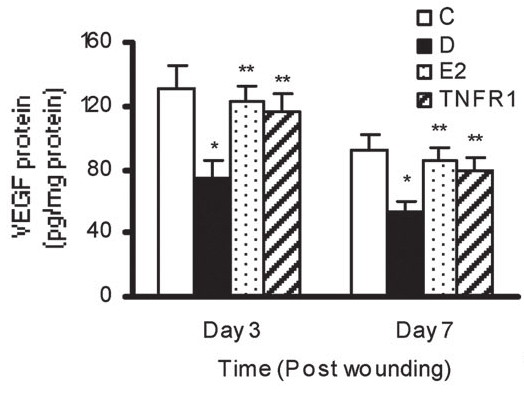}   
			\caption{Time course of VEGF in  normal and diabetic case wounds.
			Compare the simulation with the data reported in \cite[Figure 2G]{almulla2011},
			representing VEGF presence in control (white bars) and diabetic (black bars) rats.
In fact, \cite{almulla2011} investigates the connection between a defect in resolution of inflammation and the impairment of TGF-$\beta$ signaling, resulting in delayed wound healing in diabetic female rats. 
The abbreviations in the legend stand for: C, control; D, diabetic; +E2, diabetic-treated with estrogen; TNFR1, diabetic treated with the TNF receptor antagonist PEG-sTNF-RI; VEGF, vascular endothelial growth factor. }
      \label{fig:sim-diab-V}
\end{figure}

The VEGF level during wound healing is lower in diabetic patients \cite{almulla2011,mirza2011}.
In fact, as described below, a key target for the design of new therapies has been increasing VEGF levels.
The simulation output and a biological dataset are compared in Figure \ref{fig:sim-diab-V}.

\subsubsection*{LECs \& Capillaries}
\begin{figure}[h]  
      \centering
      \includegraphics[height=4.3cm]{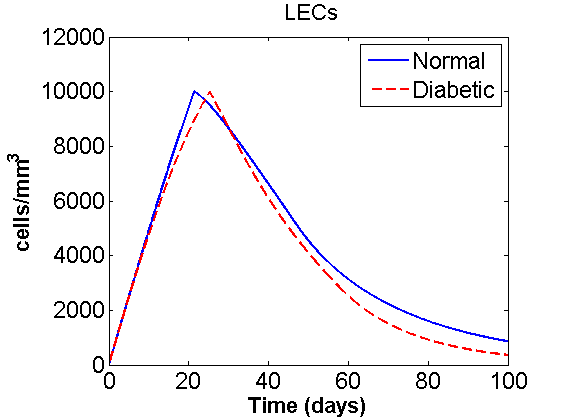}
      \includegraphics[height=4.3cm]{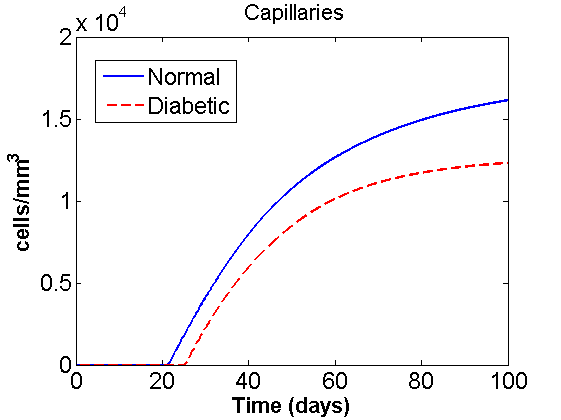}
      \includegraphics[height=4cm]{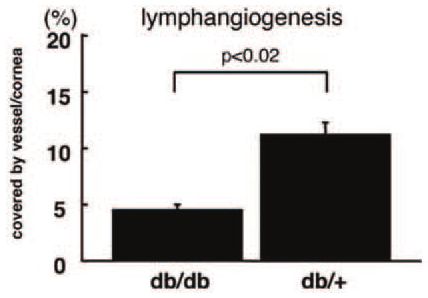}
      \caption{Time course of LEC and capillary density in normal and diabetic cases,
      compared with \cite[Figure 2b]{maruyama2007}.
The study presented in \cite{maruyama2007} investigates the role of wound-associated lymphatic vessels in corneal inflammation and in a skin wound model of wild-type and diabetic mice. The figure shows a quantification of lymphangiogenesis in the corneal suture model assay in the wild-type (db/+) and diabetic (db/db) cases. No suitable data were found in the skin wound model.
}
      \label{fig:sim-diab-L+C}
\end{figure}

In diabetic patients, lymphatic capillary formation is delayed and insufficient \cite{asai2012,maruyama2007,saaristo2006}.
The model simulations are consistent with this (Figure \ref{fig:sim-diab-L+C}).

\subsection{Analysis of the model}

In this section the steady states of the system are identified and a sensitivity analysis of the model parameters is performed.

\subsubsection{Steady States}

For the parameter set studied, at $t=0$ there are no LECs in the wound, but subsequently they increase towards a positive value of approximately $2\times 10^5$ cells/mm$^3$. However, when they reach the ``threshold'' density $L^*=10^4$ cells/mm$^3$, the system steady states change and $L$ starts to decrease towards zero. In the meantime, lymphatic capillaries start forming; their final value will depend on the dynamics of the system, but in any case it will be bigger than $C^*=10^4$ cells/mm$^3$.
This ``switch'' in the steady state values is due to the presence of two piecewise defined functions ($\sigma$ and $f$) in the system.
On the other hand, there is always one stable steady state for $M$ which also defines one for $T$:
$$
M^{eq}\approx 1875 \mbox{ cells/mm}^3 \quad 
       \mbox{and} \quad T^{eq}=\frac{a_M}{d_1}(T_L+r_1M^{eq})M^{eq} \approx 30 \mbox{ pg/mm}^3 \; .
$$
Note that the $M$ steady state (and thus also that for $T$) is unique for parameters with biologically relevant values.
For the $V$-equilibrium, the following expression is found:
$$
V^{eq}=\frac{s_V+r_3 M^{eq}}{d_3+\gamma L^{eq}} \; .
$$
Note that $V^{eq}$ depends on $L$; therefore $V$ will tend to different values according to the current $L$-steady state; for $L^{eq}=0$ it is $V^{eq}=0.5$ pg/mm$^3$.
Details about how these steady states were determined can be found in \ref{appSS}.

\begin{figure}[p]  
      \centering
      \includegraphics[height=4.1cm]{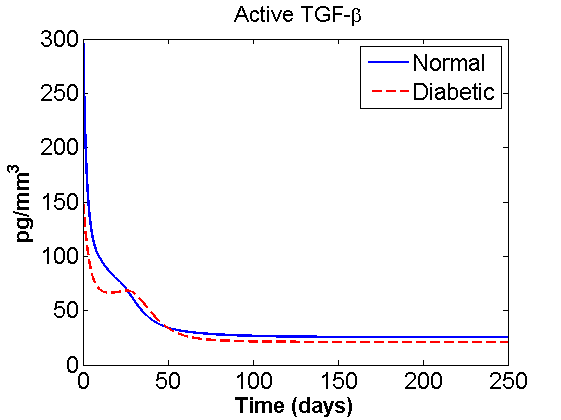}
      \includegraphics[height=4.1cm]{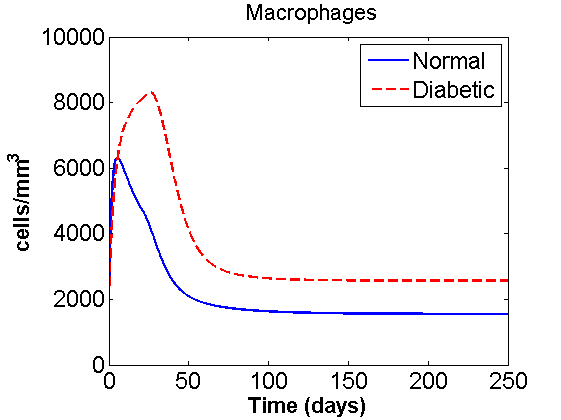}
      \includegraphics[height=4.1cm]{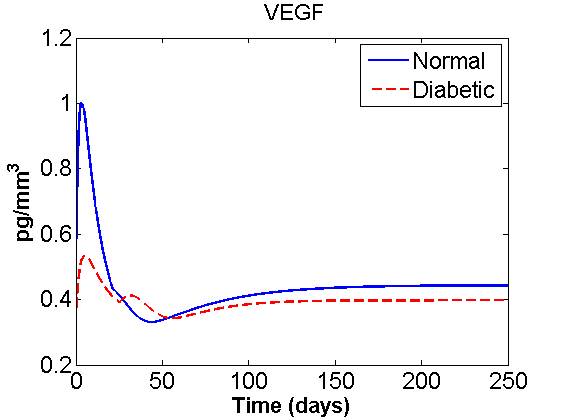}
      \includegraphics[height=4.1cm]{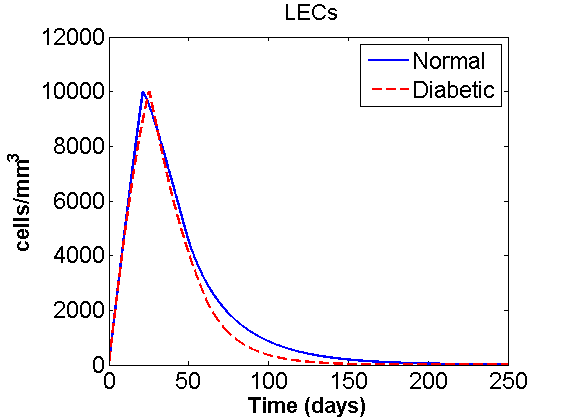}
      \includegraphics[height=4.1cm]{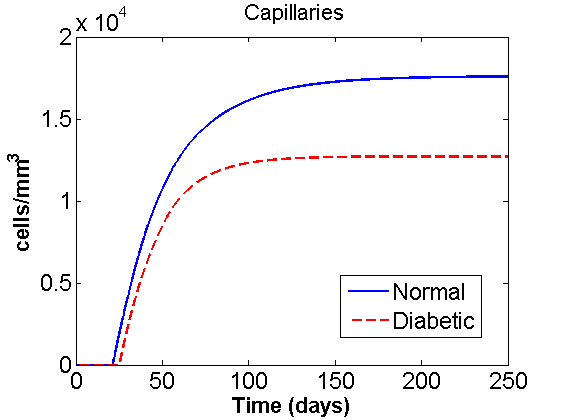}
      \caption{ Simulation of the model in both normal (solid blue) and diabetic (dashed red) cases over a time period of 250 days. }
      \label{fig:sim250}
\end{figure}

The stability of the steady states is determined numerically. The stability of $M^{eq}$ is deduced from the shape of the numerically-plotted $M$-nullcline, and the stability of the other steady states can be inferred from the simulations of the full system. See for instance the simulation shown in Figure \ref{fig:sim250}, where the model is run over a time interval of 250 days: here it is evident that all the variables tend to stay at a stable value after about 100 days post-wounding. 
Since some of the parameters were modified to simulate diabetes-related conditions, the steady states for the diabetic case are different than the corresponding ``normal'' ones. In particular, TGF-$\beta$, VEGF and capillary equilibrium values are lower in the diabetic case, while the macrophage level is higher than in the normal case. LECs go to zero in both cases.  

Although this analysis does not give very profound insights into the understanding of lymphangiogenesis, it provides some extra information about the dynamics of the system. More specifically, it shows that for realistic parameter values the system has only one steady state, which is in agreement with experimental observations.

\subsubsection{Parameter Sensitivity Analysis}

Here a numerical parameter sensitivity analysis of the model is presented which plays two important roles. On the one hand, it demonstrates which parameters are most significant in the model,
and thereby provides a deeper understanding of the dynamics of the system. On the other hand, it constitutes the first step towards the design of new therapeutic approaches.

\begin{figure}[h]  
\makebox[\textwidth][c]{\includegraphics[width=17cm]{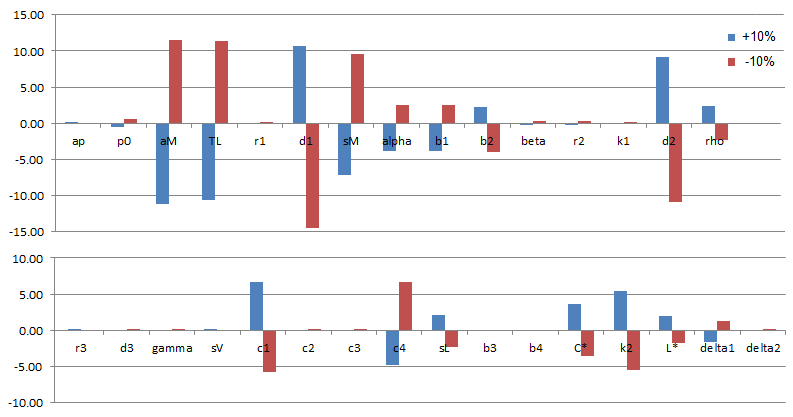}}
\caption{Percentage change in the final capillary density $C(100)$ when every parameter is increased/decreased by 10\%.}
      \label{fig:PSA}
\end{figure}

To estimate the dependence of the model on a given parameter $p$, 
a quantification of the affect of a change in $p$ on the (final) capillary density $C$ at day 100 is calculated.
To begin, $p$ is increased by 10\% and thereafter the system is solved over the time interval $[0,100]$. The final value of the capillary density thus obtained, denoted $C^{+10\%}$, is then compared with the reference value $C^{ref}$ of the corresponding density in the original system. 
The percentage change is defined by
\begin{equation}  \label{eq:PercentChange}
\mbox{ percentage change in }C = 100 \times \frac{C^{+10\%}-C^{ref}}{C^{ref}}  \quad  .
\end{equation}
The same procedure is then repeated, this time substituting the parameter $p$ with its value \emph{decreased} by 10\% and the corresponding change $C^{-10\%}$ is calculated.
The results are summarised in Figure \ref{fig:PSA}.

It is notable that perturbing any parameter does not result in a percentage change in final capillary density of more than 15\%.
Thus, the model is quite robust in terms of dependence on the parameters. Percentage changes over 5\% are observed  only for eight parameters.
Of these, one needs to \emph{decrease} $a_M$, $T_L$, $s_M$ or $c_4$ to observe an increase in the final capillary density;
while a similar enhancement is obtained by \emph{increasing} $d_1$, $d_2$, $c_1$ or $k_2$.


\section{Therapies}  \label{sec:therapies}

\subsection{Existing experimental treatments}

Although there is at present no approved therapy for enhancement of lymphangiogenesis 
(in wound healing or in any other context), many studies and experiments have been published exploring potential treatments.
In the following three such experiments are reported and then simulated.


\subsubsection*{Administration of TGF-$\beta$ Receptor-Inhibitor}
This substance binds to the TGF-$\beta$ receptors on the surface of surrounding cells,
thus making them ``insensitive'' to TGF-$\beta$ molecules and their effect.
\cite{oka2008} reports a study of the effect of TGF-$\beta$ on lymphangiogenesis in which human LECs are cultured and quantified after treatments with TGF-$\beta$1 or T$\beta$R-I inhibitor to assess cell growth, cord formation and cell migration.
It is observed that TGF-$\beta$1 treatment decreases cord formation, while T$\beta$R-I inhibitor treatment increases it.
These results are consistent with those found in \cite{clavin2008}, where it is reported that a
higher level of TGF-$\beta$1 is associated with delayed recruitment and decreased proliferation of LECs during wound repair.

\begin{figure}[h]  
    \centering
      \makebox[\textwidth][c]{\includegraphics[height=4.2cm]{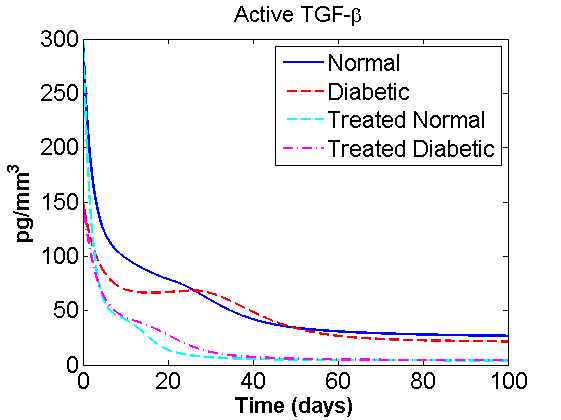} 
      \includegraphics[height=4.2cm]{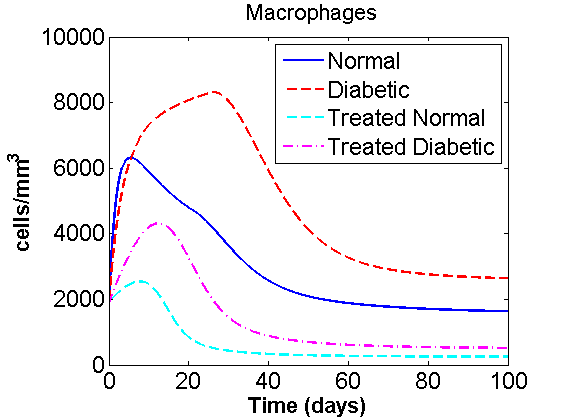}
      \includegraphics[height=4.2cm]{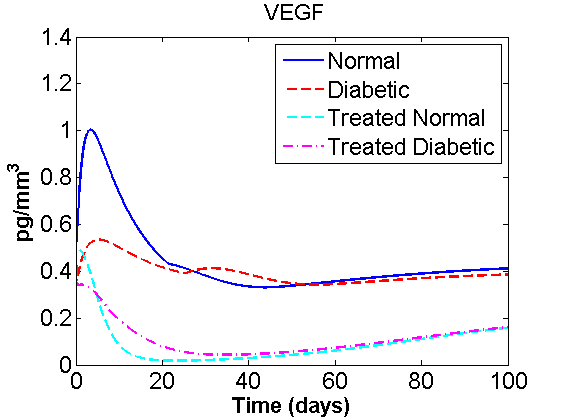}}
      \makebox[\textwidth][c]{\includegraphics[height=4.2cm]{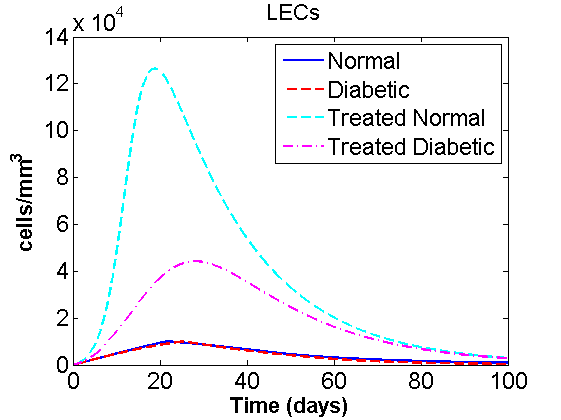} 
  		\includegraphics[height=4.2cm]{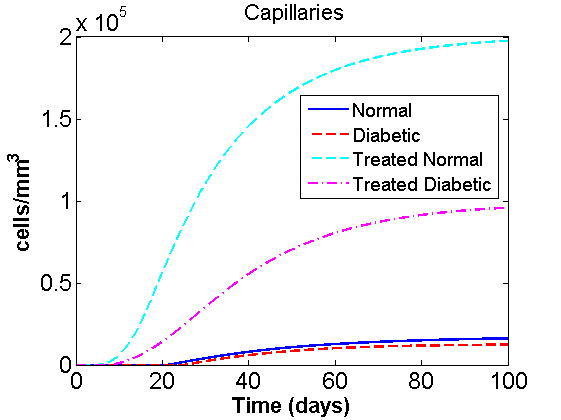} 
			\includegraphics[height=4.2cm]{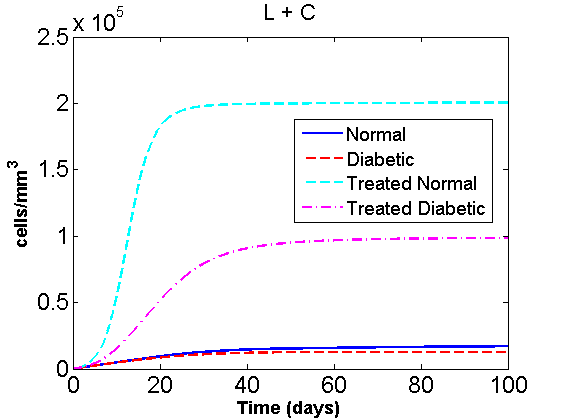}}
    \caption{ Time courses of $T$, $M$, $V$, $L$, $C$ and $L+C$ in a simulation of the T$\beta$R-I inhibitor experiment described in \cite{oka2008}. }
		\label{fig:oka}
\end{figure}

To simulate the treatment with T$\beta$R-I inhibitor, the cell migration assay is considered.
Here, the inhibitor was added at 3 $\mu$M = 817 pg/mm$^3$ (the molecular weight is 272).
Since this is significantly bigger than the concentration of TGF-$\beta$ in our model and in normal skin 
(in both cases the maximum level is 300 pg/mm$^3$),
this treatment is simulated by setting the parameters $\alpha$ and $c_4$ equal to zero 
(that is, TGF-$\beta$ molecules have no effect on cells because their receptors are ``occupied'' by the inhibitor).
The effect of this ``virtual treatment'' are shown in Figure \ref{fig:oka}, and match well with the described TGF-$\beta$ inhibitor experiment: LEC and capillary densities are markedly increased by the treatment.

\subsubsection*{Macrophage-based treatments}
Another therapeutic approach is to add macrophages to the wound, 
so that they secrete VEGF and other substances that are known to induce lymphangiogenesis. 
In \cite{kataru2009} an ``opposite'' experiment is implemented:
here a systemic depletion of macrophages is reported to markedly reduce lymphangiogenesis.
This is in accordance with \cite{watari2008}, in which it is observed that 
the induction of macrophage apoptosis inhibits IL-1$\beta$-induced lymphangiogenesis.
One hypothesis suggested to explain such results is that
because of the reduced level of macrophages, less VEGF is produced and this impairs LEC proliferation and capillary formation.

\begin{figure}[h]  
    \centering
      \makebox[\textwidth][c]{\includegraphics[height=4.2cm]{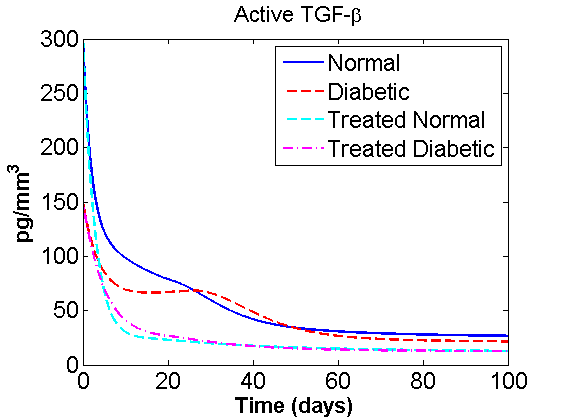} 
      \includegraphics[height=4.2cm]{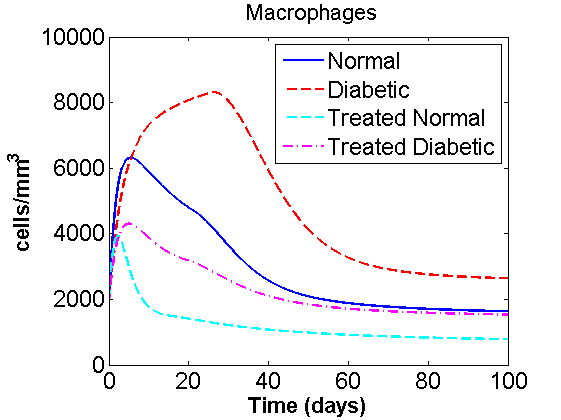}
      \includegraphics[height=4.2cm]{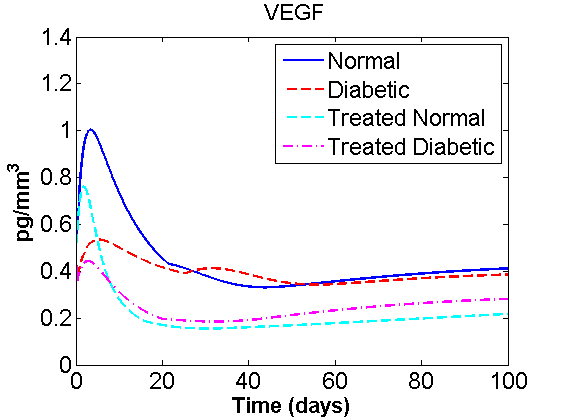}}
     \makebox[\textwidth][c]{ \includegraphics[height=4.2cm]{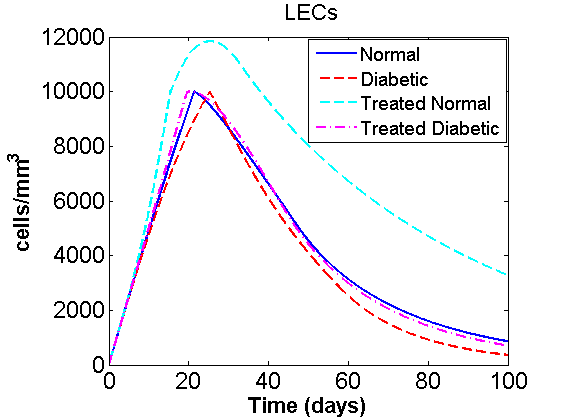} 
  		\includegraphics[height=4.2cm]{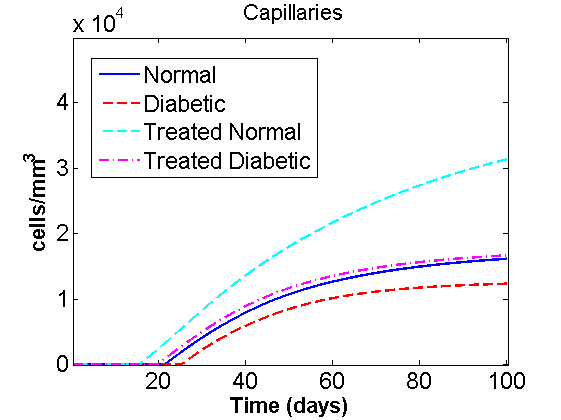} 
			\includegraphics[height=4.2cm]{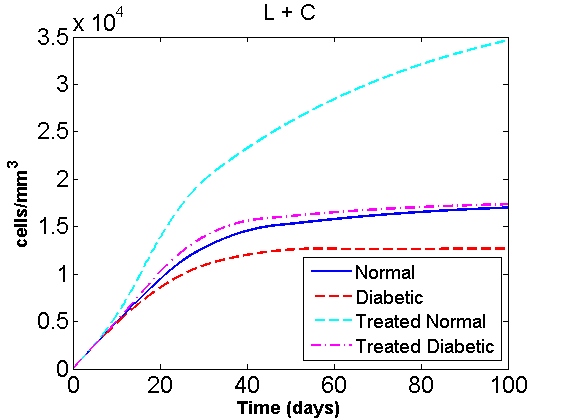}}
    \caption{ Time courses of $T$, $M$, $V$, $L$, $C$ and $L+C$ in a simulation of the macrophage-depletion experiment
		described in \cite{kataru2009}. }
		\label{fig:kataru}
\end{figure}

We simulated the increase in macrophage apoptosis by taking a bigger (for instance, the double) value of $d_2$ in the system.
The output of the model in which $d_2$ is doubled (both in normal and diabetic cases) 
is reported in Figure \ref{fig:kataru}.
In this case, the output is in contrast with what is described in the biological studies:
although fewer macrophages and consequently less VEGF are present,
more LECs and capillaries form after the simulated treatment.

\begin{figure}[h]  
    \centering
      \makebox[\textwidth][c]{\includegraphics[height=4.2cm]{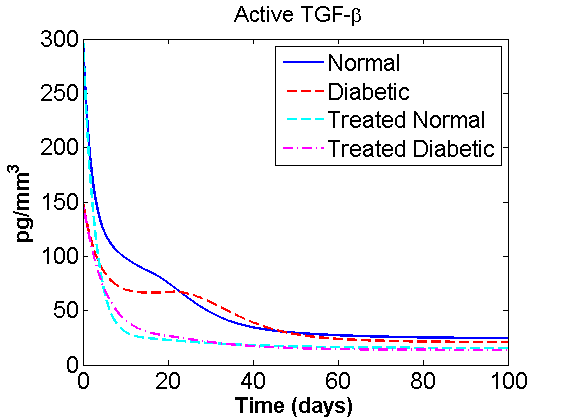} 
      \includegraphics[height=4.2cm]{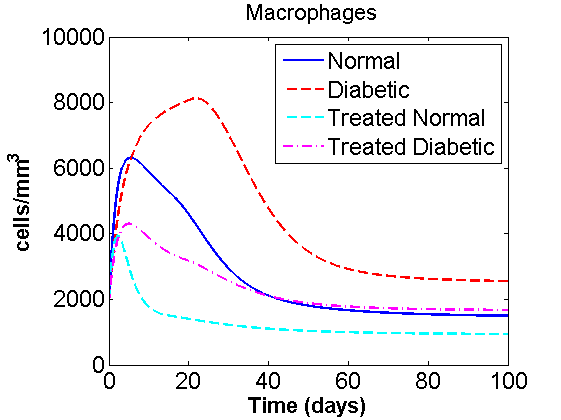}
      \includegraphics[height=4.2cm]{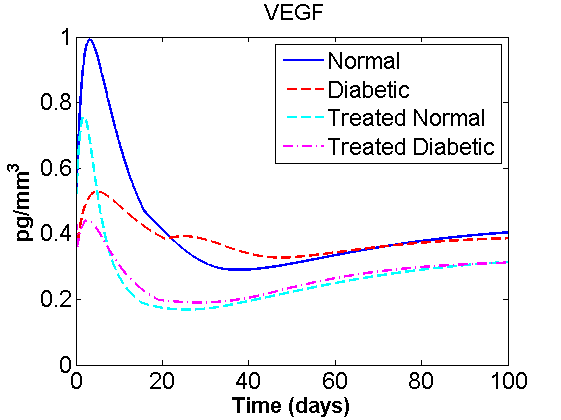}}
      \makebox[\textwidth][c]{\includegraphics[height=4.2cm]{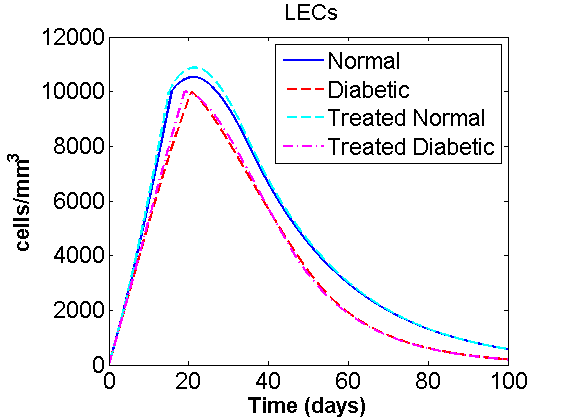} 
  		\includegraphics[height=4.2cm]{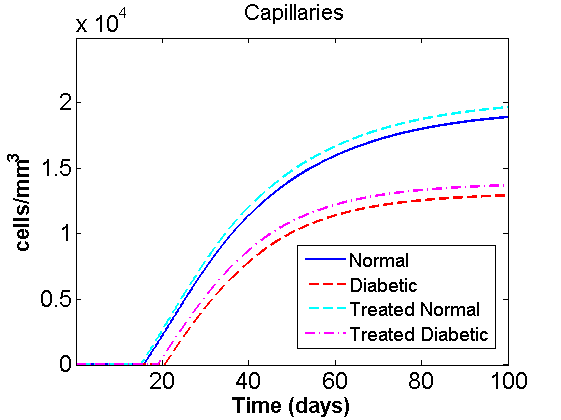} 
			\includegraphics[height=4.2cm]{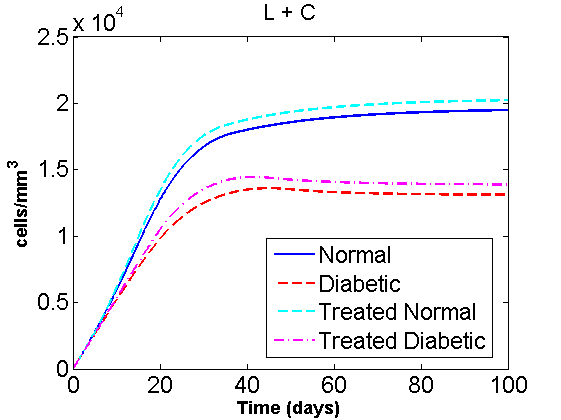}}
    \caption{ Time courses of $T$, $M$, $V$, $L$, $C$ and $L+C$ in a simulation of the macrophage-depletion experiment described in \cite{kataru2009}, where the $T$ in the LEC growth inhibition term is substituted by $T^{eq} = 30$ pg/mm$^3$. }
		\label{fig:macrophagetreatT=30}
\end{figure}

This result could be explained by the fact that, in the model, a reduction in macrophage density implies a reduction in TGF-$\beta$ level, so that the inhibition of LEC proliferation is smaller and hence there are more endothelial cells to form the capillaries. In fact, in the previous section it was found that the system is much more sensitive to $c_4$ 
than to $c_2$, $c_3$ or $\delta_2$.
It is then natural to consider the effect of fixing $T=30$ pg/mm$^3$ in the LEC growth term (\ref{eq:Leq-growth}); this level of $T$ corresponds to the TGF-$\beta$ equilibrium.
The simulation output in this case is shown in Figure \ref{fig:macrophagetreatT=30}.

With $T$ fixed, the difference between the treated and untreated cases is very small, 
but still an increase in capillary formation is observed in spite of the lower VEGF level. 
This may be due to the fact that, with fewer macrophages, the crowding term in the LEC equation is smaller, which facilitates the growth and accumulation of endothelial cells.
In fact, if the $M$ in the crowding term is fixed at its equilibrium value of 1,875 cells/mm$^3$,
there is no difference at all between treated and untreated cases.
Note that this result could have been foreseen from the parameter sensitivity analysis (Figure \ref{fig:PSA})
which predicted that a 10\% increase in $d_2$ induces a 5 to 10\% \emph{increase} in final capillary density.


\subsubsection*{VEGF supply}
A third documented approach to enhance lymphangiogenesis consists of supplying VEGF to the wound, 
since this protein promotes both LEC growth and the ability of LECs to form a network-like structure.
For instance, in \cite{zheng2007} a wound healing assessment is done in normal and diabetic mice
after a VEGF-treatment.
More precisely, two different types of VEGF were studied: VEGF-A$_{164}$ and VEGF-E$_{NZ7}$.
The authors observed that the treatment with VEGF-A$_{164}$ increased macrophage numbers and the extent of lymphangiogenesis in both wild-type and diabetic cases, while VEGF-E$_{NZ7}$ does not induce any significant change.

\begin{figure}[h]  
    \centering
      \makebox[\textwidth][c]{\includegraphics[height=4.2cm]{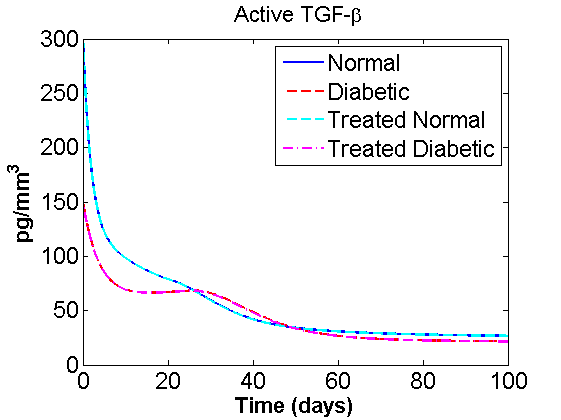} 
      \includegraphics[height=4.2cm]{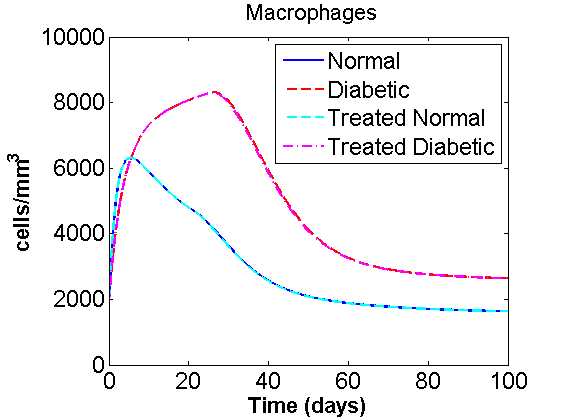}
      \includegraphics[height=4.2cm]{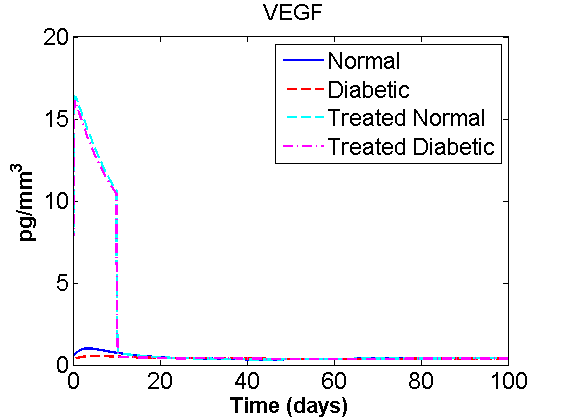}}
      \makebox[\textwidth][c]{\includegraphics[height=4.2cm]{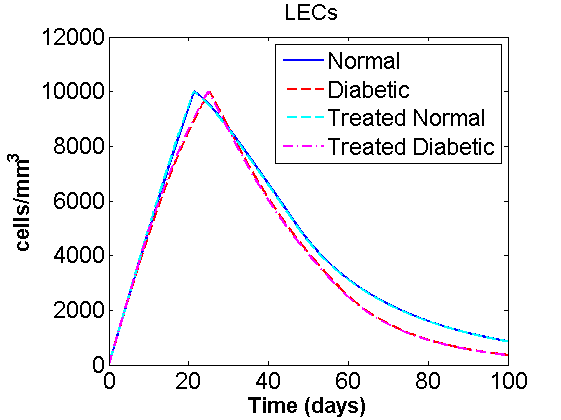}  
  		\includegraphics[height=4.2cm]{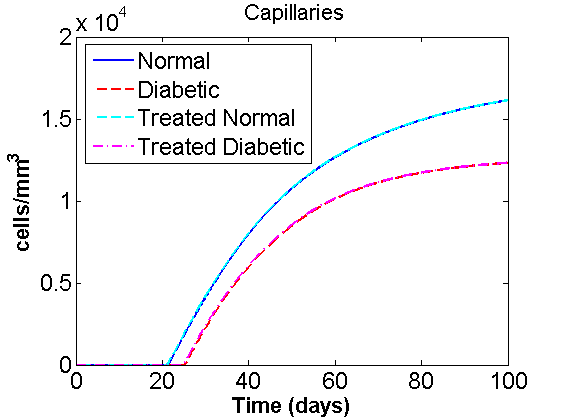} }
    \caption{ Time courses of TGF-$\beta$, macrophage, VEGF, LEC and capillary densities in a simulation of the 10-days VEGF-supply experiment described in \cite{zheng2007}. Here the original model is altered by adding a constant VEGF supply of $1.8\times 10^2$ pg/mm$^3$ for $0\leq t \leq 10$. }
		\label{fig:zheng}
\end{figure}

In order to reproduce the experiment \emph{in silico}, 
the amount of supplied VEGF is estimated in \ref{app-VEGFsupply}.
Then, a 10 days constant VEGF supply of $1.8\times 10^2$ pg/mm$^3$ is introduced in the model system.
The output is reported in Figure \ref{fig:zheng}.

\begin{figure}[h]  
    \centering
      \makebox[\textwidth][c]{\includegraphics[height=4.2cm]{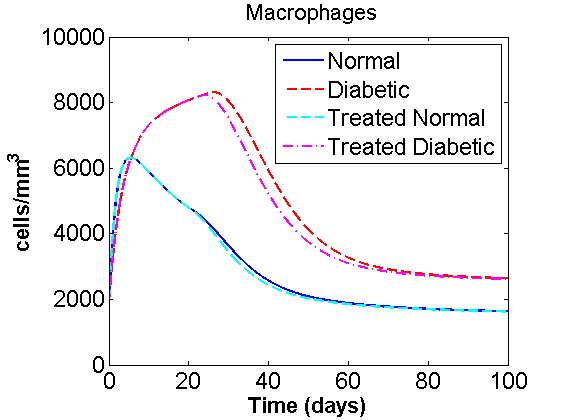}
      \includegraphics[height=4.2cm]{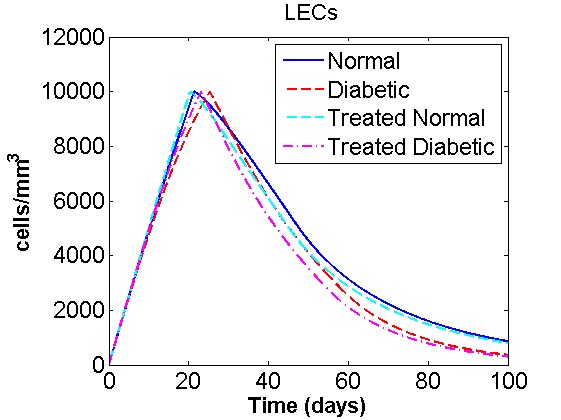} 
  		\includegraphics[height=4.2cm]{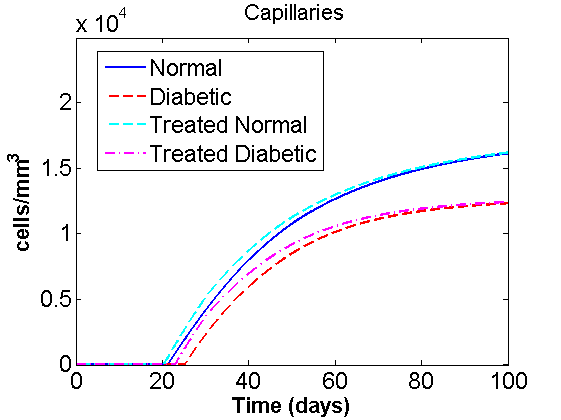}} 
		\caption{ Time courses of $M$, $L$, $C$ and $L+C$ in a simulation of the 30-days VEGF-supply experiment.}
		\label{fig:vegftreat20}
\end{figure}

There is apparently no difference between capillary formations of treated and untreated cases.
Moreover this result is relatively insensitive to the amount of VEGF supplied.
What if the same treatment is applied for 30 days instead of 10?
A simulation of this is shown in Figure~\ref{fig:vegftreat20}.

There is now a clear difference in the treated cases (especially the diabetic one) showing a lower level of macrophages and an earlier onset of capillary formation, even if the final capillary density is similar to that in the untreated case.

\subsubsection*{Observation}
There is an important feature common to all three modelled therapies:
in order to stimulate lymphatic capillary formation
one cannot consider TGF-$\beta$ or VEGF levels individually.
A precise balance of TGF-$\beta$ and VEGF is necessary for successful lymphangiogenesis.
This mutual equilibrium may be reached \emph{in vivo} by the production of either of these growth factors
by other cell types not considered in the present work. In particular, this suggests the model does not take into
account certain elements of the process.
Nevertheless, the model does effectively describe both normal and diabetic lymphangiogenesis in wound healing
which suggests that the variables considered here are the most relevant.
This indicates that potential therapies should focus on these aspects of the regeneration process.


\subsection{Novel therapeutic approaches}

As mentioned above, parameter sensitivity analysis proves useful in designing novel therapeutic approaches.
Among the ``sensitive'' parameters only $a_M$, $d_2$ and $c_1$ vary between the normal and diabetic cases
(note that the \emph{increasing $d_2$}-case was discussed above in the macrophage-based treatment).
Thus, at least theoretically, $a_M$, $d_2$ and $c_1$ are the natural targets for a therapeutic strategy aiming to increase the final lymphatic capillary density. The feasibility of each suggested parameter change is now explored.

\subsubsection*{Decreasing $a_M$} Decreasing $a_M$ means lowering the macrophage-mediated activation of TGF-$\beta$.
First of all, note that the increase in final capillary density due to a decrease in $a_M$ is explained by the fact that less active TGF-$\beta$ implies less TGF-$\beta$-inhibition of LEC growth and hence a larger LEC growth term.
For the practical implementation of this change, it is recalled that receptor-mediated TGF-$\beta$ activation consists of the binding of Latency Associated Peptide to the cell surface through receptors such as TSP-1/CD36, M6PR and multiple $\alpha$V-containing integrins \cite{taylor2009}.
Hence, a decrease of $a_M$ might be obtainable by blocking these receptors.

\subsubsection*{Increasing $c_1$} Increasing $c_1$ would be achieved by increasing the LEC growth rate.
Several possible implementations of this are found in the literature.
\emph{Recombinant human IL-8} induces (human umbilical vein  and dermal microvascular) endothelial cell proliferation and capillary tube organization \cite{li2003}.
\emph{DNA dependent protein kinase} (DNA-PK) is well known for its importance in repairing DNA double strand breaks; 
in \cite{mannell2010} it is observed that DNA-PKcs suppression induces basal endothelial cell proliferation.
In \cite{luo2013} it is reported that \emph{polydopamine}-modified surfaces were beneficial to the proliferation of endothelial cells.
Finally, \emph{non-thermal dielectric barrier discharge plasma} is being developed for a wide range of medical applications, including wound healing; in particular, \cite{kalghatgi2010} reports that endothelial cells treated with plasma for 30s demonstrated twice as much proliferation as untreated cells, five days after plasma treatment.

\subsubsection*{Other parameters}
To increase the final capillary density, one could also think about targeting other parameters to which the system is sensitive.
In particular:
\begin{itemize}

\item \emph{Decreasing $T_L$} means reducing available (latent) TGF-$\beta$ 
and hence reducing the TGF-$\beta$-inhibition over LECs.
Suppression of TGF-$\beta$ by antibodies has been proposed as a possible therapy to reduce scar formation \cite{eslami2009,finnson2013,shah1995}.
Thus many studies of TGF-$\beta$ antibodies are available.

\item \emph{Decreasing $c_4$} involves reducing the (inhibitory) effect of TGF-$\beta$ on LECs,
which can be achieved by blocking specific TGF-$\beta$-receptors on the endothelial cell surface.
Now, TGF-$\beta$ signalling is very well studied \cite{mullen2011};
in particular, it is known that TGF-$\beta$ family proteins act through two type II and two type I receptors and that ALK-1 antagonizes the activities of the canonical TGF-$\beta$ type I receptor, T$\beta$RI/ALK-5, in the control of endothelial function \cite{derynck2013}.
Moreover, a few studies have been published which deal with blocking of TGF-$\beta$ receptors in the specific case of endothelial cells \cite{liao2011,meeteren2012,watabe2003}.

\item Changes in the other ``sensitive'' parameters do not appear feasible.
Increasing $d_1$ would mean increasing the TGF-$\beta$ decay rate;
decreasing $s_M$ would mean reducing the constant source of macrophages;
increasing $k_2$ requires an increased ``carrying capacity'' for the wound.
We are not aware of practical approaches that could cause these changes.

\item Finally, among the parameters that, when changed by 10\% of their value, induce a change in final capillary density between 2 and 5\% (that is, a bit less than those analysed above),
only $b_1$ merits discussion.
\emph{Reducing $b_1$} corresponds to reducing macrophage chemotaxis towards TGF-$\beta$,
which might be achievable by blocking specific receptors on the macrophage surface.

\end{itemize}


\section{Conclusions}  \label{sec:conclusions}

Our model procures new insights into the mechanisms behind lymphangiogenesis in wound healing.
The major contributors to the process have been identified (TGF-$\beta$, macrophages, VEGF and LECs); 
the self-organisation hypothesis for the lymphatic network formation described in \cite{boardman2003,rutkowski2006} has been confirmed
and the importance of the \emph{balance} between different factors has been highlighted.
Moreover, the present work suggests novel therapeutic approaches to enhance the lymphangiogenic process in impaired wound healing.
In addition, nearly all of the relevant parameters have been estimated from biological data  and therefore this work provides fairly reliable numerical values for the parameters encountered.
However, any parameter estimation is limited by, for example, the specific experimental method used or discrepancies between the system considered here and that studied in a given reference.  The results should therefore be viewed with care. In particular, the numerical values pertaining to the aforementioned balance between the TGF-$\beta$ and VEGF may be shifted under an alteration of the parameter set.

This paper is intended as a first step in studying wound healing lymphangiogenesis through mathematical modelling.
Further work should include a spatial variable (and thus involve PDEs rather than ODEs) in order to take into account the important role of lymph flow in lymphatic capillary network formation. Introducing a spatial variable would also enable a fuller description of chemotaxis.
A PDE model would also be able to reflect further differences between angiogenesis and lymphangiogenesis. In particular, contrary to blood angiogenesis, lymphangiogenesis is unidirectional: as opposed to sprouting from both sides of the wound,
LECs appear to predominantly migrate downstream to the wound space in the direction of the interstitial flow \cite{boardman2003}.
The model could also be extended to include other aspects of wound healing, such as blood angiogenesis: implemented effectively,
this would give a more detailed overview of the different mechanisms and the time-scales involved in the various processes.

\section*{Acknowledgment}

A.B. would like to give special thanks to Jonathan Hickman for carefully proofreading this document and making numerous suggestions which improved the final presentation.



\appendix

\section{Parameter estimation}\label{appPAR}

\subsection{Equilibrium values and standard sizes}

\subsubsection*{TGF-$\beta$ equilibrium $T^{eq}$} 

The equilibrium value of active TGF-$\beta$ is about 30 pg/mm$^3$ \cite[Figure 2]{yang1999}.

\subsubsection*{Macrophage equilibrium $M^{eq}$}

The macrophage steady state can be estimated from \cite[Figure 1]{weber1990}, 
which plots typical macrophage density in the skin.
This shows that there is an average of about 15 macrophages per 0.1mm$^2$ field.
Assuming a visual depth of 80 $\mu$m, the macrophage density becomes
15 cells/(0.1mm$^2\times 0.08$mm) = 1875 cells/mm$^3$.

\subsubsection*{VEGF equilibrium $V^{eq}$}

The VEGF equilibrium concentration is estimated to be 0.5 pg/mm$^3$ from \cite[Figure 1]{hormbrey2003} and \cite[Figure 2]{papaioannou2009}.

\subsubsection*{Final LEC and Capillary density}

In \cite{rutkowski2006} we find that ``it was not until \emph{day 60}, when functional and continuous lymphatic capillaries appeared normal'' and ``at \emph{day 60} the regenerated region had a complete lymphatic vasculature, the morphology of which appeared similar to that of native vessels''.
Hence, if we assume that a capillary network that can be considered ``final'' appears at day 60, 
we will take $C^{fin}$ (or $C^{eq}$) to be the number of LECs present at this time,
and since in the normal case \emph{all} the lymphatic endothelial cells will become part of the capillary network,
we will further assume that $L^{eq}=0$.
In \cite[Figure 2E]{rutkowski2006} we see that at that time there are about 80 cells. 
This value corresponds to a 12 $\mu$m thin section.
In addition, through \cite[Figure 2D]{rutkowski2006} we can calculate the observed wound area, which is about $5.6\times 10^5 \, \mu\mbox{m}^2$.
In this way we get a volume of 0.0067 mm$^3$ with 80 cells, which corresponds to $C^{fin}=1.2\times 10^4$ cells/mm$^3$.

\subsubsection*{EC size and weight}

\cite{haas1997} reports the cross-sectional area of an EC as $10\mu\mbox{m}\times 100 \, \mu\mbox{m}$.
In \cite{vandenberg2003} the thickness of these cells is given as $0.5 \mu\mbox{m}$.
Hence we can assume a cell volume of approximately $500\mu\mbox{m}^3 = 5\times 10^{-7}$ mm$^3$.
Moreover, if we assume the density of the cells to be 1 g/mL (the same as that of water), 
we have that a cell weighs about $5\times 10^{-10} \mbox{ g} = 500 \mbox{ pg}$.

\subsubsection*{VEGF molecular weight}

The molecular weight of VEGF is $40 \mbox{kDa}=66.4\times 10^{-9}$ pg/mol \cite{roskoski2007}.


\subsection{TGF-$\beta$ equation}

\subsubsection*{Enzyme-mediated activation rate $a_p$}

For $a_p$, it seems reasonable to take the rate at which LAP binds to the receptors.
Now, in \cite{decrescenzo2001} we find an estimate for the binding rate to be about $1.7\times 10^4\mbox{ M}^{-1}\mbox{s}^{-1}$ \cite[Tables I and IV]{decrescenzo2001}.
Considering a protein weight of approximately 50 kDa (found in the same article) and converting the units we find:
$a_p \approx 2\times 10^{-5}$ mm$^3$pg$^{-1}$min$^{-1} = 2.9\times 10^{-2}$ mm$^3$pg$^{-1}$day$^{-1}$.

\subsubsection*{Receptor-mediated activation rate $a_M$}

\cite{nunes1995} reports that (``activated'') macrophages plated at $2\times 10^5$ cells/well in a 24-well tissue culture dish 
(that is, about 10$^2$ cells/mm$^3$) activated approximately 8\% of the total TGF-$\beta$ secreted after 22 hours.
This means that 1 cell/mm$^3$ activated about 0.087\% of TGF-$\beta$ per day.
On the other hand, in \cite{gosiewska1999} we find that macrophages cultured at $1\times 10^6$ cells/dish in 100-mm plates (that is, about 10 cells/mm$^3$) activated 12.2\% of total TGF-$\beta$1 after about 36 hours.
So 1 cell/mm$^3$ activated approximately 0.8\% of the TGF-$\beta$ per~day.
Hence, for $a_M$ we will take a value between 0.087 and 0.8,
say the average 0.45.

\subsubsection*{TGF-$\beta$ production rate $r_1$}

In \cite{khalil1993} it is reported that 10$^6$ macrophages produced about 30 pg of TGF-$\beta$ after 24 hours.
So one single macrophage produced $30\times 10^{-6}$ pg/day of TGF-$\beta$, 
and we can then take $r_1 = 3\times 10^{-5} \mbox{ pg}\cdot\mbox{cells}^{-1}\cdot\mbox{day}^{-1}$.

\subsubsection*{TGF-$\beta$ decay rate $d_1$}

In \cite{kaminska2005} it is stated that ``free TGF-$\beta$ has a half life of about 2 min''. We will therefore take the decay rate $d_1$ of active TGF-$\beta$ to be 
$ d_1 = {\ln 2}/{(2\mbox{ min})} = 0.35 \mbox{ min}^{-1} = 500\mbox{ day}^{-1}$.

\subsubsection*{Constant amount of latent TGF-$\beta$ $T_L$}

At the equilibrium, the TGF-$\beta$ equation becomes: $[ 0 + a_M M^{eq} ] \cdot [ T_L + r_1 M^{eq} ] - d_1 T^{eq} = 0$ .
Substituting the values of the parameters $a_M$, $r_1$, $d_1$ and of the equilibrium values $T^{eq}$ and $M^{eq}$ found before,
we get an equation for $T_L$. Solving it, we find $T_L = 18.0916$ pg/mm$^3$.
To compare this value with a ``real'' one, we consider \cite{oi2004}.
Here, taking an average of 6 pg of TGF-$\beta$1 per mg of skin (from \cite[Figure 3]{oi2004}) and 
assuming a skin density of 1 g/mL (as for water), we have a concentration of latent TGF-$\beta$ of 6 pg/mm$^3$,
which is of the same order as our previous estimate.


\subsection{Macrophage equation}

\subsubsection*{Fraction of monocytes migrating into the wound that differentiate into macrophages~$\alpha$}

We follow \cite{waugh2006} and take $\alpha$ to be equal to 0.5 in normal wound healing 
(reflecting the fact that in this case the number of inflammatory macrophages is the same of the repair ones),
and $\alpha = 0.8$ in a diabetic wound 
(since there are more inflammatory than repair macrophages this time).
However, it must be noted that \cite{waugh2006} comments that 
``there is currently no quantitative data on which the value of $\alpha$ for diabetic wounds can be based''.

\subsubsection*{Migration of monocytes to the wound in response to TGF-$\beta$: shape of $h_1(T)$ and value of $b_2$}
\begin{figure}[h]  
      \centering
      \includegraphics[height=6cm]{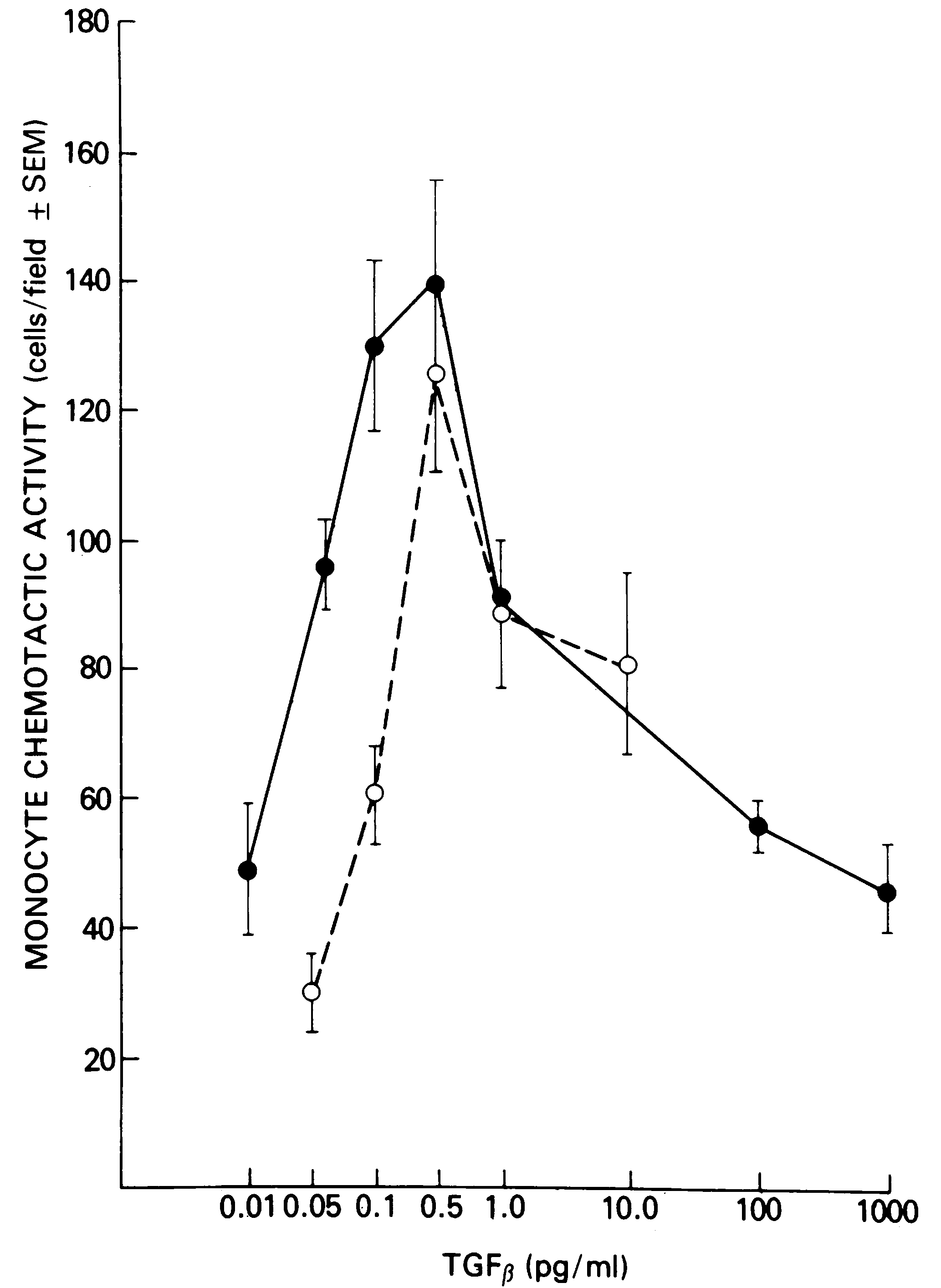}
            \caption{ Experimental data from \cite[Figure 1]{wahl1987} reporting a quantification of monocyte chemotaxis for different concentrations of TGF-$\beta$.
Chemotactic activity is defined as the mean number of monocytes that migrated through the 5-$\mu$m pores in three standard fields for each of triplicate filters. }
      \label{fig:wahldata}
\end{figure}

In \cite{wahl1987} the authors study the migration of monocytes taken from healthy volunteers and observe that the motility of the cells depends on the dose of TGF-$\beta$ to which they are exposed. 
Moreover, even if the response varied with individual donors, they see that
``the optimal chemotactic concentration for TGF-$\beta$ fell within the range 0.1-1.0 pg/mL''.
These findings are shown in Figure \ref{fig:wahldata}, which shows the dataset reported in \cite{wahl1987}.
In light of these observations, it is reasonable to take the chemotactic function $h_1(T)$ to be
$$ h_1(T) = \frac{ b_1 T^m}{( b_2 + T^{2m})} $$
where we will take $m=2$ from visual comparison with data. 

To determine $b_2$ we look for the maximum of $h_1(T)$.
This is located at the point $T_{max}$ where $h_1'(T_{max})=0$, which is $T_{max}=\sqrt[2m]{b_2}$.
\cite{yang1999} reports that the normal level of TGF-$\beta$ in the skin is about 30 pg/mm$^3$,
and that this amount increases up to 300 pg/mm$^3$ during wound healing.
From this, and from the observation that macrophage levels also reach a peak soon after this TGF-$\beta$ peak \cite{nor2005},
we deduce that the maximum monocyte/macrophage migration occurs when the level of TGF-$\beta$ in the skin is about 300 pg/mm$^3$.
Thus we take $b_2 = 300^{2m}$; for $m=2$ this gives $b_2 = 300^4 = 8.1\times 10^9 \; \mbox{pg}^4 \; (\mbox{mm}^{-3})^4$.
This seems to be in contrast with the data reported in \cite[Figure 1]{wahl1987}: here $T_{max}$ is around 0.5 pg/mL = $0.5\times 10^{-3}$ pg/mm$^3$. However, this value does not seem realistic and should be considered carefully. In particular, we recall that chemotaxis occurs through \emph{gradients} of a chemical, and considering this directed movement to depend only on the absolute concentration of the chemoattractant is a simplification. 
In fact, in the experiment described in \cite{wahl1987} the diluted chemotactic stimuli were placed in the bottom wells of microchamber plates that were separated from the upper wells by a filter with 5.0 $\mu$m pores; then monocytes were put in the plates with human TGF-$\beta$ diluted at different concentrations. Chemotactic activity was defined as the mean number of monocytes that migrated through the pores.
Therefore, we consider the value of $T_{max}$ to be a ``conventional'' one and we keep the estimate for $b_2$ found above.

\subsubsection*{Percentage of monocytes/macrophages undergoing mitosis $\beta$}

In \cite[Figure 4]{greenwood1973} we see that only about 0.5\% of monocytes \emph{in vitro} show mitotic activity.
Therefore we take $\beta = 0.005$.

\subsubsection*{Macrophage growth rate $r_2$ and carrying capacity $k_1$}

To obtain estimates for these parameters, we first consider just the logistic part of the macrophage equation 
${dM}/{dt}=r_2M(1-M/k_1)$, whose solution is 
\begin{equation}  \label{eq:MacrGrowToFit}
M(t) = \frac{k_1 M_0 e^{r_2 t}}{k_1 + M_0 (e^{r_2 t}-1)} \; .
\end{equation}
In \cite{zhuang1997} murine macrophage-like cell growth is measured under different conditions. We then fit the data contained in \cite[Figure 1]{zhuang1997} to the curve (\ref{eq:MacrGrowToFit})
using the MatLab function \texttt{nlinfit}.
Moreover, \cite{zhuang1997} specifies that 
``$6\times 10^6$ cells were cultured in 100-mm tissue culture plates in 10 mL of the (above) medium''.
Then, taking $M_0 = 600 \mbox{ cells/mm}^3$, 
we get the estimates $\widehat{r_2} = 1.22$, $\widehat{k_1} = 6\times 10^5$,
with 95\% confidence intervals 
$(0.82,1.62)$ and $(4.40\times 10^5 , 7.61\times 10^5)$ respectively
 (these are calculated using the MatLab function \texttt{nlparci}).

\subsubsection*{Macrophage constant removal rate $d_2$}

\cite{cobbold2000} presents a mathematical model for keloid and hypertrophic scarring. Here we find that macrophages are known to exist in a wound for several days after the initial migration; so based on this we assume a decay rate for macrophage cells to be of the order of $d_2 \approx 0.2 \mbox{ day}^{-1}$.

\subsubsection*{Macrophage capillary-dependent removal rate $\rho$}

We assume that the term $\rho C$ becomes of the same order of $d_2$ when capillaries reach their ``final'' density, 
which we have estimated as $1.2\times 10^4$. Since we estimate $d_2 = 0.2$, we take $\rho = 10^{-5}$.

\subsubsection*{Migration of monocytes to the wound in response to TGF-$\beta$: $b_1$}

To find $b_1$ we notice that this parameter determines the maximum level of macrophages $M$ during healing.
To find this value, we refer to \cite{nor2005}. 
Here the authors investigate the role of TGF-$\beta$ activation in wound repair, and assess several components of wound healing (including inflammatory cell infiltration) over a period of 28 days.
\cite{nor2005} reports that at day 5 a maximum of about 70 macrophages/field (400x) are observed.
Assuming a diameter field of view of 0.4 mm and a depth of field of 80 $\mu$m, 
1 field corresponds to $(0.2)^2\pi \mbox{ mm}^2 \times 80 \times 10^{-3} \mbox{mm} \approx 10^{-2}\mbox{mm}^3$.
Then 70 cells/field $\approx$ 7000 cells/mm$^3$.
Numerical experimentation shows that reproduction of this result requires $b_1 \approx 8\times 10^8$ cells mm pg$^2$/day.

\subsubsection*{Macrophage constant source $s_M$}

At the steady state, the $M$-equation becomes
$$
s_M + \alpha \frac{ b_1 T^2}{b_2 + T^4} + \beta r_2 M \left( 1-\frac{M}{k_1} \right) - d_2 M  - \rho C M = 0 \; .
$$
Substituting the equilibrium values and the parameters found above, we get 
$$ s_M = 586 - b_1 \times 5.5 \times 10^{-8} \; . $$
Above, we chose $b_1 = 8\times 10^8$ cells mm pg$^2$/day.
Therefore $s_M = 542$ cells/day.


\subsection{VEGF equation}

\subsubsection*{VEGF production by macrophages $r_3$}

From \cite[Figure 1B]{kiriakidis2003} we have that human macrophages plated at $10^6\, \mbox{cells}/\mbox{ml} = 10^{3}\, \mbox{cells}/\mbox{mm}^3$ produced $214 \mbox{ pg}/\mbox{ml} = 214\times 10^{-3} \mbox{pg}/\mbox{mm}^3$ of VEGF after 24 hours of culture.
Then $r_3 \approx 8.9\times 10^{-6}\mbox{pg}\cdot\mbox{cells}^{-1}\cdot\mbox{h}^{-1} 
          = 2.1 \times 10^{-4} \mbox{ pg}\cdot\mbox{cells}^{-1}\cdot\mbox{day}^{-1}$.
In fact, this value is a bit smaller than the one we use in the model,
which we take to be $9\times 2.1 \times 10^{-4} \approx 1.9 \times 10^{-3}$ pg/cell/day.
This ``adjustment'' is done considering the data shown in \cite[Figure 2]{sheikh2000}, which reports that the VEGF peak (occurring at day 5) corresponds to a level of about 1000 pg/mL = 1 pg/mm$^3$.
Since in our model we assume that the VEGF peak is due (mainly) to the macrophages,
it seems reasonable to adjust the parameter $r_3$ to meet this observation.

Other estimates were obtained from \cite{xiong1998,zhang1997} and one ``equivalent'' parameter was found in the modelling paper \cite{gabhann2007}.
Although the numerical values are different, they are all between $10^{-6}$ and $10^{-4}$ pg/cell/day. This variety is not surprising because different cell types produce VEGF at different rates (as clearly shown in \cite{zhang1997}).
Since in the context of wound healing lymphangiogenesis we are mainly concerned with macrophages, we focus more on the values for these cells \cite{kiriakidis2003,xiong1998}.

\subsubsection*{VEGF decay rate $d_3$}

The half-life for $\mbox{VEGF}_{\footnotesize{165}}$ (the most common and biologically active VEGF protein) at room temperature is 90 minutes \cite{kleinheinz2010}. 
It follows that $d_3 = 11 \mbox{ day}^{-1}$.      
To compare this value with those used in other modelling articles,
we see that in \cite{plank2004} the VEGF decay rate is taken to be $\mu_v = 0.456 \mbox{ h}^{-1}=10.9\mbox{ day}^{-1}$,
while in \cite{zheng2013} the VEGF natural decay/neutralization rate is $\mu_c = 0.65 \mbox{ h}^{-1}=15.6\mbox{ day}^{-1}$.

\subsubsection*{VEGF consumption/internalization by ECs $\gamma$}
\begin{figure}[h]
\begin{subfigure}[h]{0.3\textwidth}
\centering
 \includegraphics[height=4.5cm]{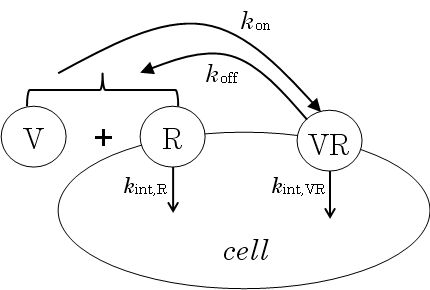}
\end{subfigure}
\hspace{3cm}
\begin{subfigure}[h]{0.3\textwidth}
\centering
\begin{tabular}{r}
		{\scshape \footnotesize parameter values}  \\
		\hline
		\footnotesize $k_{\mbox{on},V,R1} = 3.8 \times 10^6 \mbox{M}^{-1}\mbox{s}^{-1}$            \\
		\footnotesize $k_{\mbox{on},V,R2} = 1.2 \times 10^6 \mbox{M}^{-1}\mbox{s}^{-1}$            \\
		\footnotesize $k_{\mbox{off},V,R1} = 95 \times 10^{-6} \mbox{s}^{-1}$                      \\
		\footnotesize $k_{\mbox{off},V,R2} = 410 \times 10^{-6} \mbox{s}^{-1}$                     \\
		\footnotesize $k_{\mbox{int},R1} = k_{\mbox{int},R2} = 1 \times 10^{-5} \mbox{s}^{-1}$     \\
		\footnotesize $k_{\mbox{int},VR1} = k_{\mbox{int},VR2} = 28 \times 10^{-5} \mbox{s}^{-1}$  \\
		\hline
 \end{tabular}
\end{subfigure}
\caption{A schematic representation of the dynamics considered in \cite{gabhann2004}.
         $V$ stands for VEGF, $R$ for Receptor and $VR$ for the ligand-receptor complex. }
\label{fig:VRinternScheme}
\end{figure}

In \cite{gabhann2004} the VEGF internalization by a cell is described in the way schematised in Figure \ref{fig:VRinternScheme}.
In that figure and the following text, $V$ stays for VEGF, $R$ for receptor and $VR$ for the ligand-receptor complex.
     
\noindent
We will not distinguish between the two kinds of receptors $R1$ and $R2$, 
so we will just take the average of the above parameters. That is, we will take
\begin{equation}  \label{def:param}
\begin{array}{rclcrcl}
k_{\mbox{\footnotesize{on}}} &=& 2.5\times 10^6 \mbox{M}^{-1}\mbox{s}^{-1},&\quad& k_{\mbox{\footnotesize{off}}} &=& 2.5 \times 10^{-4} \mbox{s}^{-1}, \\ 
k_{\footnotesize{\mbox{int},R}} &=&  10^{-5} \mbox{s}^{-1}, &\quad& k_{\footnotesize{\mbox{int},VR}} &=& 2.8 \times 10^{-4} \mbox{s}^{-1}.
\end{array}
\end{equation}
According to this reaction scheme, the corresponding equation for $V$ (for a single cell) is
\begin{equation}   \label{eq:V-all-cells}
\frac{dV}{dt} = - k_{\mbox{\footnotesize{on}}} \cdot V \cdot R + k_{\mbox{\footnotesize{off}}} \cdot (VR) \; .
\end{equation}
Here the dimensions of $R$ are moles per unit volume. Similarly the equation for the ligand-receptor complex $(VR)$ (per cell) is
$$
\frac{d(VR)}{dt} = + k_{\mbox{\footnotesize{on}}}\cdot V\cdot R - k_{\mbox{\footnotesize{off}}} \cdot (VR) - k_{\footnotesize{\mbox{int},VR}} \cdot (VR) \; .
$$
So, at equilibrium:
\begin{equation}   \label{eq:Veq}
\begin{split}
k_{\mbox{\footnotesize{on}}}\cdot V\cdot R - k_{\mbox{\footnotesize{off}}} \cdot (VR)_{eq} - k_{\footnotesize{\mbox{int},VR}} \cdot (VR)_{eq} = 0
\\  \Rightarrow \quad (VR)_{eq} = \frac{k_{\mbox{\footnotesize{on}}}\cdot R}{k_{\mbox{\footnotesize{off}}}+k_{\footnotesize{\mbox{int},VR}}}\cdot V \quad .
\end{split}
\end{equation}

\noindent
To determine the value of $R$, we note that human endothelial cells (cultured \emph{in vitro}) display 1,800 VEGFR1/cell and 5,800 VEGFR2/cell \cite{imoukhuede2012}, giving a total of $7.6\times 10^3$ receptors/cell. Assuming a cell volume of 500 $\mu$m$^3$ we take $R=2.5\times 10^{-14}$ mol/mm$^3$.
Substituting this value for $R$ and replacing the parameters with the values given in (\ref{def:param}),
the equation (\ref{eq:Veq}) gives $(VR)_{eq} \approx 118 \cdot V = \widehat{VR}_{eq} \cdot V$.
Substituting this value for $(VR)$ in (\ref{eq:V-all-cells}) we have
$$
\frac{dV}{dt} = 
   - \underbrace{ \left( k_{\mbox{\footnotesize{on}}} \cdot R - k_{\mbox{\footnotesize{off}}} \cdot \widehat{VR}_{eq} \right) }
   _{ = 2.9\times 10^3 \mbox{ \footnotesize{day}}^{-1} } \cdot V \cdot L
$$
Then, again assuming a cell volume of $500 \mu\mbox{m}^3$,
we have that our parameter (per cell) is \\$\gamma = 5\times 10^{-7} \mbox{ mm}^3 \times 2.9\times 10^3 \mbox{ day}^{-1} \mbox{cells}^{-1}$
                                       = $1.4\times 10^{-3} \mbox{ mm}^3 \cdot \mbox{day}^{-1} \cdot \mbox{cells}^{-1}$.

\subsubsection*{VEGF constant source $s_V$}

At equilibrium, the $V$-equation becomes
$$
s_V + r_3 M - d_3 V - \gamma V L  = 0
$$
Substituting the equilibrium values and the parameters gives $s_V =  1.9$ cells/day.

\subsubsection*{VEGF supply in \cite{zheng2007}}  \label{app-VEGFsupply}

In \cite{zheng2007} control medium with or without growth factors was injected into wounded mice 
at a dose of 2 $\mu$g/wound/10 days.
Since the experimental wounds are full-thickness with a 5-mm diameter, 
and assuming a skin thickness of 0.56 mm \cite{hansen1984},
this amount corresponds approximately to $1.8\times 10^4$ pg/mm$^3$/day.
Now, since the experiment is performed \emph{in vivo},
it is reasonable to assume that the vast majority of the VEGF is washed away and dispersed by body fluids.
In the absence of quantitative data, we assume that only 1\% of the added VEGF is therapeutically active in the wound, giving a delivery of $1.8\times 10^2$ pg/mm$^3$/day.


\subsection{LECs equation}

We consider first the parameter $c_1$, then $k_2$, and then the remaining parameters.
For $c_1$ and $k_2$ we focus on the ``logistic'' part of the equation
  \begin{equation}   \label{LEClogistic}
  \frac{dL}{dt} = c_1 L - \frac{L^2}{k_2} = c_1 L \left( 1 - \frac{L}{c_1 k_2} \right)  \; .
  \end{equation}  
Recalling that the volume of an EC is approximately 500 $\mu$m$^3$, 
closely packed cells have a density of $1 \mbox{ cell} / 500 \mu\mbox{m}^3 = 2 \times 10^{6} \mbox{cells}/\mbox{mm}^3$.
We assume the carrying capacity to be 10\% of this, so that $c_1 k_2 \approx 2 \times 10^{5} \mbox{cells}/\mbox{mm}^3$.      
In the following, we will fit experimental data to the solution of (\ref{LEClogistic}), which is
   \begin{equation}  \label{logisticSOL}
   L(t) = \frac{c_1 k_2 L_0 e^{c_1 t}}{c_1 k_2 + L_0 (e^{c_1 t}-1)}  \; .
   \end{equation}

\subsubsection*{``Normal'' proliferation rate $c_1$}

In \cite{nguyen2007} the different responses of lymphatic, venous and arterial endothelial cells to angiopoietins is studied; note that the cells were isolated and cultured from bovine mesenteric vessels.
\cite[Figure 4B]{nguyen2007} shows the evolution of LEC density in time.
After converting these data (in particular, those corresponding to control, or 10\% FBS) into cells/$\mbox{mm}^3$,
we can fit the function (\ref{logisticSOL}) to them, 
obtaining the estimate $\widehat{c_1}=0.42\mbox{ day}^{-1}$
with 95\% confidence interval $(0.15 , 0.70)$.
Again, we can compare this value with other estimates obtained from different biological sources, such as \cite{muller1987,tsai2006,whitehurst2006}; also, a similar parameter is estimated in \cite{zheng2013}.
Although the numerical values for $c_1$ found in these other references are all different, it is noticeable that they are all around $10^{-1}$. This make us very confident in estimating this parameter.

\subsubsection*{Maximum density of cells (per unit time) $k_2$}

In the previous section we found different possible values for $c_1$.
Since $c_1 k_2 = 2 \times 10^{5} \mbox{cells}/\mbox{mm}^3$, we can easily obtain $k_2$:
\begin{itemize}
\item From \cite{nguyen2007} we get $k_2\approx 4.71 \times 10^5 \mbox{ cells}\cdot\mbox{day}/\mbox{mm}^3$ (this is the value used in our model); 
\item From \cite{muller1987} we get $k_2\approx 4.77 \times 10^5 \mbox{ cells}\cdot\mbox{day}/\mbox{mm}^3$ for bovine cornea
      \\and $k_2 \approx  5.43 \times 10^5 \mbox{ cells}\cdot\mbox{day}/\mbox{mm}^3$ for bovine fetal heart;
\item From \cite{tsai2006} we get $k_2 \approx 3.13 \times 10^5 \mbox{ cells}\cdot\mbox{day}/\mbox{mm}^3$;
\item From \cite{whitehurst2006} we get $k_2 \approx 4.48 \times 10^5 \mbox{ cells}\cdot\mbox{day}/\mbox{mm}^3$ for 10\% FBS 
      \\and $k_2 \approx 18.52 \times 10^5 \mbox{ cells}\cdot\mbox{day}/\mbox{mm}^3$ for 2\% FBS.
\end{itemize}

\subsubsection*{VEGF-dependence of LECs growth $c_2,c_3$}
    
		To estimate $c_2$ and $c_3$, we consider only the exponential VEGF-dependent part of the LEC equation, that is
      \begin{equation}  \label{VEGFdep}
      \frac{dL}{dt} = \left( c_1 + \frac{V}{c_2 + c_3 V} \right) \cdot L   \;  ,
      \end{equation}
      whose solution is
      \begin{equation}  \label{VEGFdepSOL}
      L(t) = L_0 \cdot \exp \left[ \left( c_1 + \frac{V}{c_2+c_3V} \right)\cdot t \; \right]  \; .
      \end{equation}
      Recall that we already have an estimate for $c_1$. Also, notice that, if time is fixed, (\ref{VEGFdepSOL}) can be seen as a function of $V$ only.

\cite[Figure 7A]{whitehurst2006} shows the response of rat mesenteric LECs to VEGF-A and VEGF-C at low serum conditions (2\% FBS). 
The cells were seeded at the density of 16,000 per well in 24-well plates. 
Then, VEGF$_{165}$ and mature VEGF-C (2-100 ng/mL) were added 4 hours after seeding. 
Finally, cells were counted 72 hours later.
This provides a set of data giving the cell densities for different concentrations of VEGF-A and C.
These data refer to time $t=72$ hours = 3 days, and our approach is to fit the function (\ref{VEGFdepSOL}) 
as a function of $V$ (with $t$ fixed at 3 days) to the experimental data.
Recalling that 1 ng/mL = 1 pg/$\mbox{mm}^3$, and considering a standard well of 1 mL = $10^3 \mbox{mm}^3$
(for a 24-well cell culture plate), we can convert these data into suitable units and use the MatLab function \texttt{nlinfit} to fit (\ref{VEGFdepSOL}) to them.
This gives $\widehat{c_2} = 42$ days and $\widehat{c_3} = 4.1$ pg/day/mm$^3$,
with 95\% confidence intervals $(-7.7 , 92)$ and $(2.9 , 5.3)$ respectively.

\subsubsection*{TGF$\beta$-dependence of LECs growth $c_4$}

We estimate $c_4$ using experimental data obtained in the absence of VEGF. Therefore we consider only the part of the LEC equation concerning TGF-$\beta$ regulation of cell growth:
\begin{equation}  \label{TGFbdep}
\frac{dL}{dt} = \left( \frac{c_1}{1+c_4 T} \right) L
\end{equation}
whose solution is
\begin{equation}  \label{TGFbdepSOL}
L(t) = L_0 \cdot \exp \left[ \left( \frac{c_1}{1+c_4 T} \right) \cdot t \; \right]   \; .
\end{equation}
In \cite{muller1987} the inhibitory action of TGF-$\beta$ on bovine endothelial cells is studied.
\cite[Figures 1(a) and 1(b)]{muller1987} demonstrate the growth in time of bovine cornea and fetal heart endothelial cells respectively, while \cite[Figure 1(c)]{muller1987} shows the inhibition on cell growth by TGF-$\beta$.
We can thus use these figures in different ways:
     \begin{itemize}
     \item In \cite[Figures 1(a) and 1(b)]{muller1987} the amount of TGF-$\beta$ is fixed,
     so we can consider (\ref{TGFbdepSOL}) as a function of $t$ only.       
     Recalling that previously we found the value $c_1 = 0.42\mbox{ day}^{-1}$ for \cite[Figure 1(a)]{muller1987}, 
     and $c_1 = 0.37\mbox{ day}^{-1}$ for \cite[Figure 1(b)]{muller1987},
     we can fit the function (\ref{TGFbdepSOL}) to the data corresponding to $T = 2$ ng/mL = 2 pg/mm$^3$
     to find the parameter $c_4$.
     The MatLab functions \texttt{nlinfit} and \texttt{nlparci} give us the estimate
     $\widehat{c_4} = 0.16$ and its 95\% confidence interval $(-0.25 , 0.58)$ 
     for cornea ECs, and $\widehat{c_4} = 0.22$ with 95\% confidence $(-0.50 , 0.95)$
     for fetal heart ECs.
     \item Another strategy is to use the data contained in \cite[Figure 1(c)]{muller1987}, 
     fixing the time ($t$ = 4 days) and considering (\ref{TGFbdepSOL}) as a function of $T$ only.
     For bovine cornea ECs, in this case we have $\widehat{c_4} = 0.32$ 
     with 95\% confidence interval $(0.12 , 0.51)$.
     Similarly the data for fetal heart ECs gives $\widehat{c_4} = 1.3$ 
     with 95\% confidence interval $(-0.66 , 3.2)$.
     \end{itemize}
Looking at the confidence interval for each of the estimates found above,
we argue that the most ``reliable'' values for $c_4$ are the first three: $\widehat{c_4} = 0.16$, $\widehat{c_4} = 0.22$ and $\widehat{c_4} = 0.32$.
Hence, in our model we chose to take the average of these numbers, that is $\widehat{c_4} = 0.24$ mm$^3$/pg.

To compare this number with a similar estimate found in another source,
we consider \cite[Figure 1]{sutton1991}, which also shows how cell growth is influenced by TGF-$\beta$. 
Here bovine retinal and aortic endothelial cells were plated at 25 cells/cm$^2$ and TGF$\beta$-1 was added at different proliferation stages.
Cell numbers were determined 5 days after the addition of different concentrations of TGF$\beta$-1.
Considering the data in \cite[Figure 1B]{sutton1991} and assuming a dish of 10mm height, we can fit the function (\ref{TGFbdepSOL}) to the data with the MatLab function \texttt{nlinfit},
taking $L_0 \approx 10 \mbox{ cells}/\mbox{mm}^3$ (estimated from \cite[Figure 1A]{sutton1991})
and consequently $c_1 = ({1}/{t}) \ln ({L(t)}/{L_0})=0.37\mbox{ day}^{-1}$, since $t = 5$ days. 
This gives the estimate $\widehat{c_4} = 6$ with 95\% confidence interval $(-2.5 , 14)$.

\subsubsection*{Threshold levels $L^*$,$C^*$}

In \cite{rutkowski2006} it is observed that in a wound space LECs begin to organize in a network-fashion after about 25 days.
Hence, we take $L^*$ to be the number of LECs present at day 25 during normal repair.
In \cite[Figure 2E]{rutkowski2006} we see that at that time there are about 80 cells. This value corresponds to a section of width 12 $\mu$m.
In addition, through \cite[Figure 2D]{rutkowski2006}, we can calculate the observed wound area, which results to be $5.6\times 10^5 \, \mu\mbox{m}^2$.
In this way we get a volume of 0.0067 mm$^3$ which contains 80 cells, which corresponds to $1.2\times 10^4$ cells/mm$^3$.
We take $L^*$ to be $10^4$~cells/mm$^3$.

For $C^*$, we assume that LECs stop coming into the wound when capillaries reach a level which is not far from the final one, that we have estimated to be $C^{fin}=1.2\times 10^4$ cells/mm$^3$.
Therefore we take $C^* = 10^4$~cells/mm$^3$. 
Note that our estimate for $L^*$ and $C^{fin}$ are the same because in \cite[Figure 2E]{rutkowski2006} it happens that the cell numbers counted at day 25 and at day 60 are about the same.


\section{Calculation of the steady states} \label{appSS}

For $T$ we have immediately
\begin{equation}  \label{eq:Tequil}
T^{eq} = \frac{a_M}{d_1} (T_L + r_1 M^{eq}) M^{eq} \;  ,
\end{equation}
since the exponential tends to zero as $t\rightarrow\infty$.
It follows that there is one $T$-steady state for every $M$-steady state.

\begin{figure}[h]
     \centering
     \includegraphics[height=6cm]{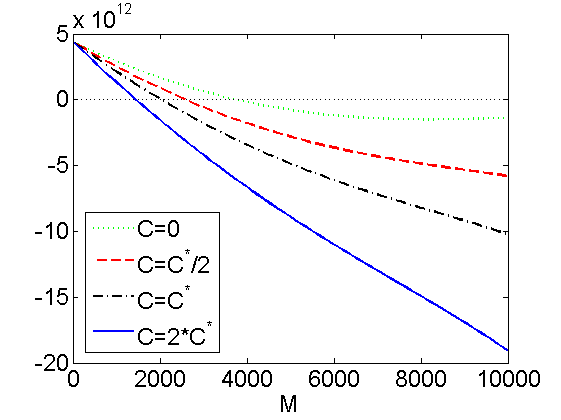}    
     \includegraphics[height=6cm]{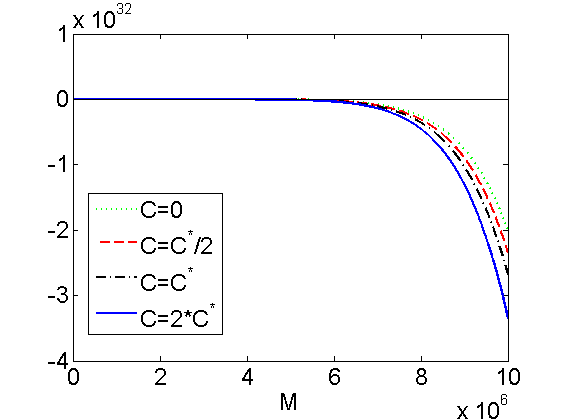} 
	\caption{Plots of the $M$-expresson at the equilibrium.
	         The different lines refer to different representative values of $C$. }  
	\label{fig:MequilpolGraph}
\end{figure}

For $M$ the situation is more complicated.
Writing down the equation (\ref{eq:Meqn}) at the equilibrium and rearranging the terms one finds
\begin{equation} \label{eq:Mequil1}
-\beta\frac{r_2}{k_1} M^2  +  (\beta r_2 - d_2 - \rho C) M  +  s_M + \alpha \frac{b_1 T^2}{b_2 + T^4}  =  0   \;  .
\end{equation}
Plugging in the expression for $T$ found in (\ref{eq:Tequil}),
the equation (\ref{eq:Mequil1}) becomes a polynomial in $M$ of degree 10.
Plotting this polynomial for $M\in[0,10^4]$ and $M\in[0,10^7]$ gives the graphs shown in Figure \ref{fig:MequilpolGraph}.
The graphs are strongly indicative that there is only one stable steady state for (at least) $M<10^7$ and this is around 2000.
This finding is in agreement with the estimate $M^{eq}=1875$ cells/mm$^3$ from the biological literature.

The equilibrium equation for $V$ is
\begin{equation}  \label{eq:Vequil}
s_V + r_3 M - d_3 V - \gamma V L = 0 \quad \Rightarrow \quad V^{eq} = \frac{s_V+r_3 M^{eq}}{d_3 +\gamma L^{eq}}  \; .
\end{equation}
So there is one $V$-equilibrium for every $M$, $L$ equilibrium.

At this point, it is more convenient to look at the capillary equation first,
and afterwards consider the LEC one.
There are two different cases to consider in the $C$-equation:
\begin{itemize}

\item if $L+C < L^*$, then $\sigma = 0$ and $dC/dt = 0$ always;

\item if $L+C \geq L^*$, then $\sigma = 1$ and the equilibrium equation becomes
      $$ (\delta_1 + \delta_2 V) L  = 0  \;  ,  $$
      whose only solution is $L=0$ (since $V$ cannot be negative).

\end{itemize}

Coming to the $L$-equation, it is necessary to consider a number of different cases,
since there are two piecewise-defined functions involved:

\begin{center}
\begin{tabular}{|c|l|l|}
\hline
CASE  &  CONDITIONS                     &  FNs VALUE         \\
\hline
I     &  $L+C\geq L^*$ and $C\leq C^*$  &  $\sigma(L,C)=1$ and $f(C)=1-C/C^*$   \\
\hline  
II    &  $L+C\geq L^*$ and $C\geq C^*$  &  $\sigma(L,C)=1$ and $f(C)=0$   \\
\hline 
III   &  $L+C < L^*$ and $C\leq C^*$    &  $\sigma(L,C)=0$ and $f(C)=1-C/C^*$   \\
\hline  
IV    &  $L+C < L^*$ and $C\geq C^*$    &  $\sigma(L,C)=0$ and $f(C)=0$   \\
\hline  
\end{tabular}
\end{center}

\noindent
Note that case IV seems to be \emph{not realistic}, since the estimates for the thresholds are $L^*\approx C^*$.

\begin{itemize}

\item[\textsc{case} I] Since we must have $L=0$ in order to have a steady state in the $C$-equation,
the equilibrium $L$-equation reduces to
$$
\left( s_L + \frac{b_3 V^2}{b_4 + V^4} \right) \left( 1 - \frac{C}{C^*} \right) = 0  \; ,
$$
which implies
$$
C^{eq} = C^*  \; .
$$

\item[\textsc{case} II] Again, we must have $L=0$ for the equilibrium in the fifth equation.
This time the steady state $L$-equation is automatically satisfied 
and therefore any value of $C$ corresponds to a steady state.

\item[\textsc{case} III] In this case, rearranging the fourth equation for $L$ one gets
\begin{equation} \label{eq:LequilCaseIII}
\begin{split}
-\frac{1}{k_2}L^2 + \left[ \left( c_1 + \frac{V}{c_2+c_3 V} \right) \left( \frac{1}{1+c_4 T} \right) 
 - \frac{M+C}{k_2}  \right] L  
\\ + \left( s_L + \frac{b_3 V^2}{b_4 + V^4} \right) \left( 1 - \frac{C}{C^*} \right) = 0
\end{split}
\end{equation}
where $V$ depends on $L$ and this dependence is given by the expression (\ref{eq:Vequil}).
Notice that if $\gamma=0$ then $V$ does not depend on $L$.
Hence, if we set $\gamma=0$ the expression (\ref{eq:LequilCaseIII}) becomes a simple quadratic equation for $L$. In order to study how the system changes as $V$ depends on $L$, we gradually increase $\gamma$ and see how the roots change. 
The function (\ref{eq:LequilCaseIII}) is therefore plotted numerically versus $L$ and the resulting graphs are reported in Figure \ref{fig:LequilpolGraph-gammas}.
These show that increasing $\gamma$ does not have a significant effect on the roots of (\ref{eq:LequilCaseIII}): the function is always concave for $0\leq L \leq 4\times 10^5$ and intercepts the horizontal axis once. Thus for $C<C^*$ (which is the case we are studying) there is only one intersection for $L>0$ at about $L=2\times 10^5$.

\begin{figure}[p]
     \centering
     \includegraphics[height=4.1cm]{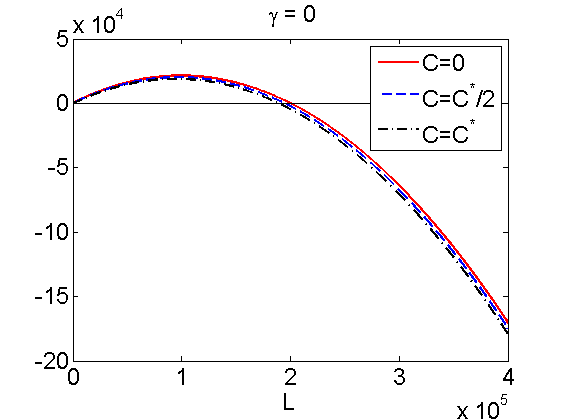}    
     \includegraphics[height=4.1cm]{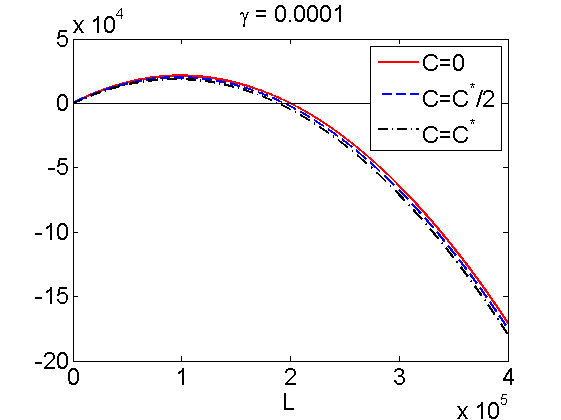} 
     \includegraphics[height=4.1cm]{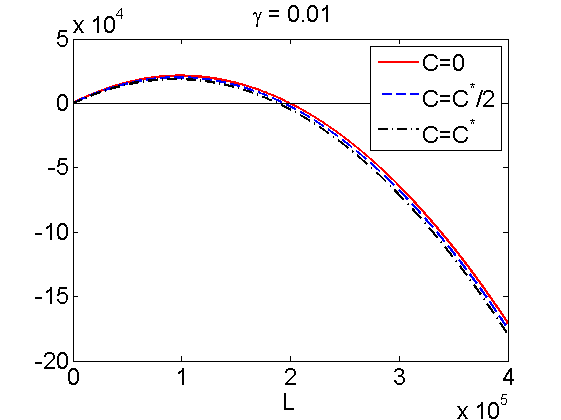} 
     \includegraphics[height=4.1cm]{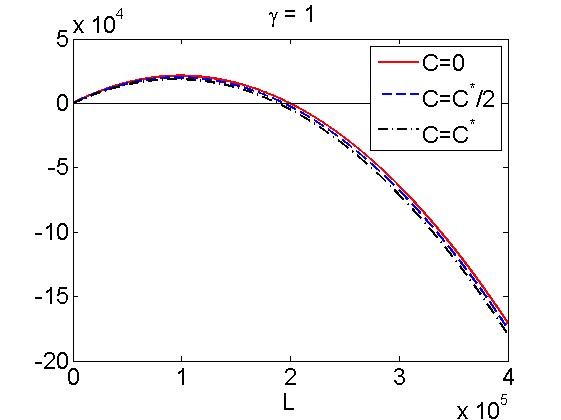} 
     \includegraphics[height=4.1cm]{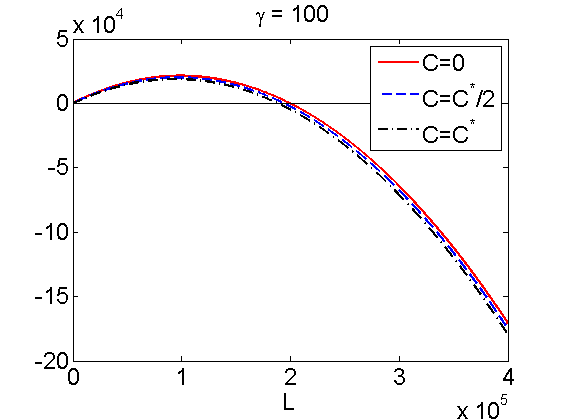} 
     \includegraphics[height=4.1cm]{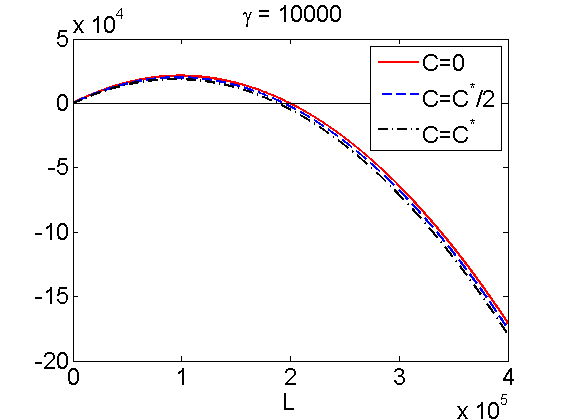} 
	\caption{Plots of the expression (\ref{eq:LequilCaseIII}) as a function of $L$ 
            for different values of~$\gamma$ 
            (more precisely, $\gamma=0, 10^{-4}, 10^{-2}, 1, 10^2, 10^4$).
	         In each picture the function is plotted 
            for three different representative values of $C$,
            in particular $0$, $C^*/2$ and~$C^*$.}  
	\label{fig:LequilpolGraph-gammas}
\end{figure}

\end{itemize}



\bibliography{mybib-bio,mybib-math}
\bibliographystyle{elsarticle-num} 

%
%
%
%

\end{document}